# Ferri-ionic Coupling in CuInP$_2$S$_6$ Nanoflakes: Polarization States and Controllable Negative Capacitance

Anna N. Morozovska[1*], Sergei V. Kalinin[2†], Eugene. A. Eliseev[3], Svitlana Kopyl[4], Yulian M. Vysochanskii[5], and Dean R. Evans[6‡]

[1] Institute of Physics, National Academy of Sciences of Ukraine, 46, pr. Nauky, 03028 Kyiv, Ukraine

[2] Department of Materials Science and Engineering, University of Tennessee, Knoxville, TN, 37996, USA

[3] Frantsevich Institute for Problems in Materials Science, National Academy of Sciences of Ukraine, Omeliana Pritsaka str., 3, Kyiv, 03142, Ukraine

[4] Department of Physics & CICECO – Aveiro Institute of Materials, Campus Universitario de Santiago, 3810-193 Aveiro, Portugal

[5] Institute of Solid-State Physics and Chemistry, Uzhhorod University, 88000 Uzhhorod, Ukraine

[6] Air Force Research Laboratory, Materials and Manufacturing Directorate, Wright-Patterson Air Force Base, Ohio, 45433, USA

## Abstract

We consider nanoflakes of van der Waals ferrielectric CuInP$_2$S$_6$ covered by an ionic surface charge and reveal the appearance of polar states with relatively high polarization ∼ 5 μC/cm$^2$ and stored free charge ∼ 10 μC/cm$^2$, which can mimic "mid-gap" states associated with a surface field-induced transfer of Cu and/or In ions in the van der Waals gap. The change in the ionic screening degree and mismatch strains induce a broad range of the transitions between paraelectric phase, antiferroelectric, ferrielectric, and ferri-ionic states in CuInP$_2$S$_6$ nanoflakes. The states' stability and/or metastability is determined by the minimum of the system free energy consisting of electrostatic energy, elastic energy, and a Landau-type four-well potential of the ferrielectric dipole polarization. The possibility to govern the transitions by strain and ionic screening can be useful for controlling the tunneling barrier in thin film devices based on CuInP$_2$S$_6$ nanoflakes. Also, we predict that the CuInP$_2$S$_6$ nanoflakes reveal features of the controllable negative capacitance effect, which make them attractive for advanced electronic devices, such as nano-capacitors and gate oxide nanomaterials with reduced heat generation.

---

[*] corresponding author, e-mail: anna.n.morozovska@gmail.com
[†] corresponding author, e-mail: sergei2@utk.edu
[‡] corresponding author, e-mail: dean.evans@afrl.af.mil

## I. INTRODUCTION

Nanosized layered van der Waals (vdW) ferroelectric materials, i.e., materials with strong in-plane covalent bonding and weak interlayer (van der Waals) interactions [1], such as $CuInP_2(S,Se)_6$ monolayers, thin films, and nanoflakes [2, 3, 4], are very interesting nanoscale objects for fundamental studies of coupling between the polar and antipolar long-range orders, space charge, and lattice strains. The free energy profile of uniaxial ferrielectric $CuInP_2(S,Se)_6$ [5, 6] has more than two potential wells [7, 8], which leads to specific features of their polarization reversal, and in particular, is responsible for screening, temperature, and strain control of the multiple energy-degenerate metastable polar states. It is important to note that the energy profiles of $CuInP_2(S,Se)_6$ can flatten in the vicinity of the nonzero polarization states [9, 10], as this is what makes the ferrielectric material different from classical ferroelectric materials with a first or second order ferroelectric-paraelectric phase transition; these potential energy profiles can be shallow or flat near the transition points, corresponding to a great number of energy-degenerate polar states. The flattened energy profiles in the vicinity of the nonzero polarization states can give rise to unusual polar, electromechanical, and dielectric properties associated with the polarization dynamics [11, 12].

Ferrielectricity in the $CuMP_2(S,Se)_6$ family (M = In or Cr) can be termed as an antiferroelectric order, but with a switchable spontaneous polarization formed by two sublattices with spontaneous dipole moments that are antiparallel and different in magnitude [13]. The order-disorder type ferrielectric transition occurs in a bulk unstrained CIPS at temperatures slightly higher (305 – 315 K) than the Curie temperature $T_C \approx$ 293 K. Below the transition temperature from the polar ferrielectric to the nonpolar paraelectric phase, the spontaneous polarization of $CuInP_2S_6$ (CIPS) is directed normal to its structural layers because of antiparallel shifts of the $Cu^+$ and $In^{3+}$ cations from the middle of the layers [14]. The $Cu^+$ cations flip in their multi-well local potential between "up" and "down" states with a temperature increase and populate these states with equal probability above the transition temperature (i.e., in the paraelectric phase). Far above the transition temperature, the potential wells can weaken or completely vanish.

In and Cu ions are located in the planes of the CIPS layers, where very large electric fields can shift them to the vdW gap. However, small out-of-plane shifts take place due to local fields. The smaller intra-layer shift of the $In^{3+}$ cations in their local potential is opposite in direction with respect to the larger intra-layer shift of the $Cu^+$ cations. These intra-layer shifts of $Cu^+$ and $In^{3+}$ cations form the ferrielectric ground state and determine the dipole length in the so-called low-polarization state with a small spontaneous polarization $P$~(1 – 4) µC/cm$^2$, whereas a "mid-gap" ferroelectric state with a high polarization $P$~(9 – 10) µC/cm$^2$ can be induced by a very large external electric field that can move the Cu$^+$ cations into the vdW gap [15]. Note that very large electric fields can also be induced by electrochemically active ions at the CIPS surface, which can stabilize the high-polarization states.

In the framework of Landau-Ginzburg-Devonshire (LGD) mean-field approximation [16, 17], the presence of Cu$^+$ and In$^{3+}$ cationic sublattices in CIPS is described by polar and antipolar order parameters, $P$ and $A$. A complete LGD thermodynamic potential describing ferrielectric CIPS with a first order phase transition contains even (2-nd, 4-th, and 6-th) powers of $P$ and $A$, as well as the biquadratic coupling between them. As shown in Ref. [17], the biquadratic coupling term $A^2 P^2$ induces a term proportional to $P^8$ in the LGD thermodynamic potential for $P$. Using the LGD potential with eighth powers of $P$, we predicted a temperature – stress phase diagram [17, 18, 19], containing the paraelectric (PE) phase and two ferrielectric states, FI2 and FI1, with smaller (~1 µC/cm$^2$) and larger (~4 µC/cm$^2$) amplitudes of the spontaneous polarization, respectively (see **Fig. 1(a)**).

In the framework of the Ising model with a mixed anisotropy of the local crystal field, [20] the polar and antipolar orders correspond to pseudo-spin projections $\vec{S} = \pm 1$ in the polar state, and $\vec{S} = 0$ in the nonpolar state. From this standpoint, the nonpolar state can be composed of three different antiferroelectric orderings, and thus it has a 3-fold degeneracy [16]. Although these states cannot be distinguished in the framework of the continuous medium LGD approach, they can be predicted by DFT calculations [21]. The "true" ferroelectric state with a high polarization is nondegenerate, corresponding to the colinear ordering of pseudo-spins.

It is well-known that the bound surface charge associated with the out-of-plane spontaneous polarization in ferroelectric thin films causes a depolarization field: this effect, which can destroy the polarization and induce a transition to the PE phase, becomes more dominant as the film thickness decreases [22]. However, it was shown from first principles that a strong polarizing field, which originates from the charged ionic layers in insulating materials, can drive paraelectric thin films into a non-centrosymmetric polar state [23]. Earlier, a phenomenological approach analyzing the behavior of ferroelectric films covered by ionic charge layers was developed by Stephenson and Highland (SH) [24, 25]. A combination of the LGD and SH approaches allows for the derivation of analytical solutions describing unusual phase states in uniaxial [26, 27, 28, 29] and multiaxial [30] ferroelectric thin films and nanoparticles [31], as well as antiferroelectric thin films with electrochemical polarization switching [32, 33]. The analysis [26-33] leads to the elucidation of ferroionic and antiferroionic states, which are the result of a nonlinear electrostatic interaction between the ionic free charges and the bound charges (ferroelectric dipoles) at the surface. The ferroionic and antiferroionic states originate from the strong nonlinear dependence of the screening charge density on the surface electric potential, also named the electrochemical overpotential [26]. The prefix "over" shows that the electric potential inside the charged layer is different from the external voltage due to the self-screening and depolarization field [34, 35]. There are plenty of mixed antiferroionic-ferroionic states characterized by pinched and/or constricted asymmetric loops, whose appearance depends on the static characteristics of surface charges (such as their concentration, formation energies, and details of the surface density of states) and their relaxation dynamics in an applied field [31].

Abovementioned works [26-33] consider the influence of ionic screening on the polar properties of bulk and nanosized ferroelectrics. To the best of our knowledge, the influence of ionic screening on the polar properties of nanosized vdW ferrielectrics is almost unexplored, while its role can be principally important. This theoretical study aims to fill the gap in knowledge and considers polar and dielectric properties of CIPS nanoflakes covered by a layer of mixed ionic-electronic nonlinear surface charge with slow relaxation dynamics in an external field. Using the LGD-SH approach, we show that transitions between PE, FI1, FI2, and ferri-ionic (FII) states take place in CIPS nanoflakes. The paper is structured as follows: **Section II** contains the formulation of the problem, description of methods, basic equations, and assumptions; **Section III** contains the results, discussion, and analysis; and **Section IV** contains a summary. Calculation details are listed in **Supplemental Materials** [36].

## II. PROBLEM DESCRIPTION

Let us consider a CIPS nanoflake, where its height $h$ is much smaller than the lateral size $2R$. The nanoflake bottom surface is clamped to a rigid conducting substrate, and a mismatch strain $u_m$ exists at the flake-substrate interface (see **Fig. 1(b)**). The axis Z is parallel to the polar axis of the nanoflake, which has a uniaxial polarization $P_3 \uparrow\uparrow z$. The nanoflake upper surface is separated from the top disk-shaped electrode (or tip apex) by the ultra-thin dielectric layer of thickness $d$. The aspect ratio of the nanoflake lateral size $2R$ to height $h$ is very large (e.g., $R \gg 10^2 h$). This assumption allows the influence of side surface features and roughness to be neglected and a cylindrical-shaped nanoflake to be considered, which provides for an analytical description.

Specifically, for the piezo-response force microscopy (PFM) geometry, the space between the nanoflake and the disk-shaped tip apex can be filled by gaseous matter with a controllable partial pressure of oxygen (or other atoms) that may be in the form of a gas, liquid, or soft matter [37]. In result, the electrochemically active species can form an ultra-thin "passive" dielectric layer at the CIPS surface [38, 39].

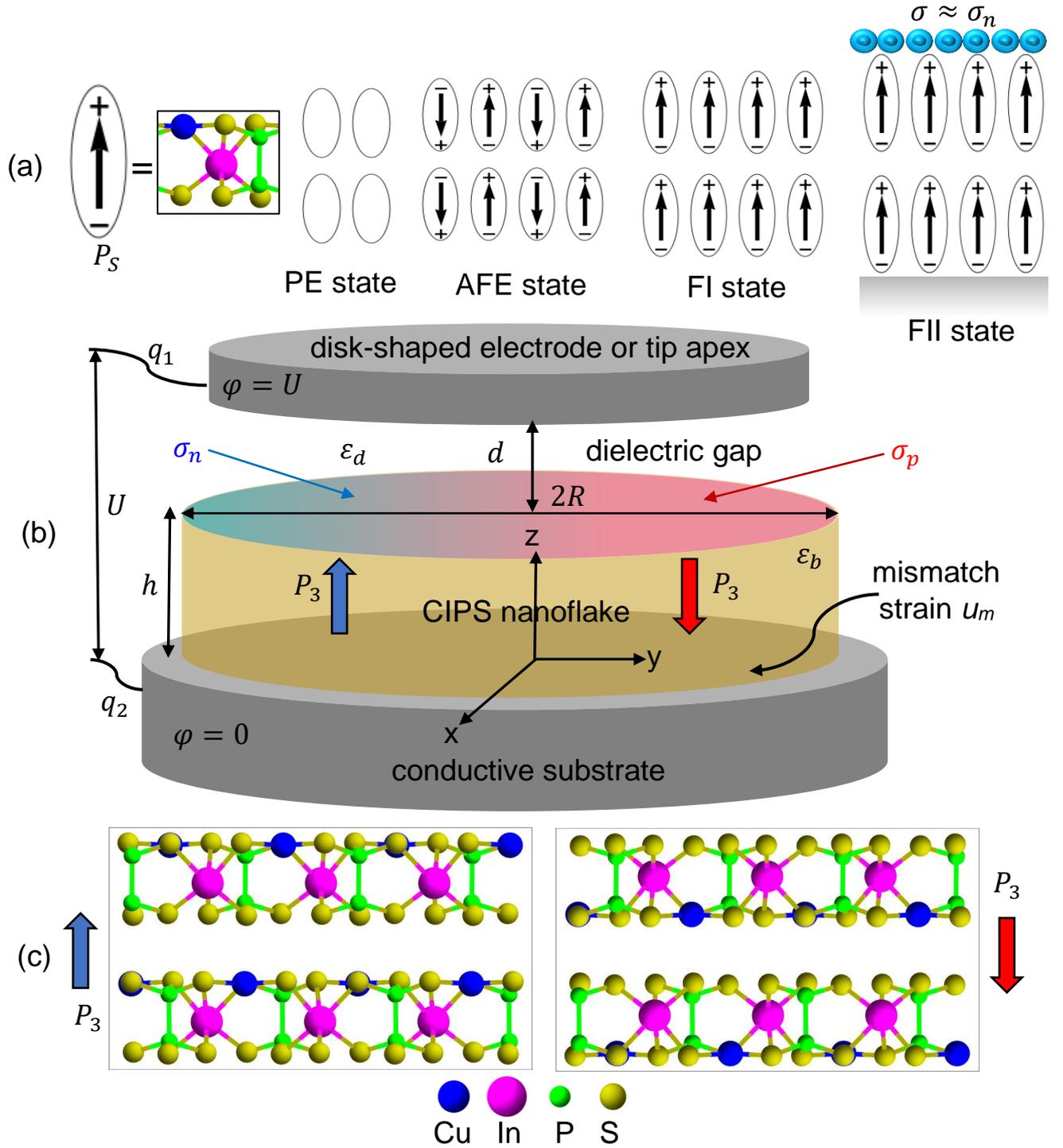

**FIGURE 1**. **(a)** Schematic illustration of the paraelectric (PE), antiferroelectric (AFE), ferrielectric (FI), and ferri-ionic (FII) states of CIPS. **(b)** A cylindrical-shaped CIPS nanoflake of height $h$ and radius $R$ placed between a top electrode and a conducting substrate. The "up" and "down" directions of the uniaxial polarization $P_3$ are shown by the blue and red arrows, respectively. The surface of the nanoflake is covered with a layer of mobile surface charges (e.g., holes, electrons, ions, vacancies, protons, hydroxyl groups, etc.), where the density $\sigma$ is the sum of positive (red color) and negative (blue color) charges, $\sigma_p$ and $\sigma_n$, respectively. $\varepsilon_b$ is the relative background permittivity of the CIPS nanoflake, $\varepsilon_d$ is the relative dielectric permittivity of the dielectric layer of thickness $d$,

and $u_m$ is the mismatch strain at the flake-substrate interface. **(c)** Atomic positions in the "up" and "down" FI states.

The relative background permittivity of the CIPS nanoflake, $\varepsilon_b$, is regarded as the same order as the dielectric permittivity of the dielectric layer, $\varepsilon_d$ ($3 < \varepsilon_{b,d} \leq 10$). The inclusion of the relative background permittivity is a common rule for a Landau-type description of the dielectric properties of various ferroics with a soft polar optical mode (i.e., for the case of ferroelectrics, ferrielectrics, and paraelectrics); this is in a full agreement with multiple experimental results proving the existence of the constant "background" permittivity far from the Curie temperature $T_C$ (see e.g., Refs. [40, 41, 42]).

Due to the electrochemically active species inside the dielectric layer, the upper surface of the CIPS nanoflake is covered with a layer of mobile surface charges (holes, electrons, ions, vacancies, protons, hydroxyl groups, etc.). The surface charge densities of positive ($\sigma_p$) and negative ($\sigma_n$) charges depend on the electric potential $\phi$ in a complex nonlinear way.

The polarization, electric field, elastic stresses, and strains in CIPS nanoflakes are calculated using the finite element modeling (FEM). In this method, we use differential equations for surface charge time dynamics along with electrostatic equations coupled with Euler-Lagrange equations for CIPS polarization obtained from a variation of the LGD free energy functional. The equations for surface charge dynamics, derived in Ref. [31], are listed in **Appendix A1** [36]. The LGD free energy functional of CIPS nanoflakes, which includes a Landau expansion of 2-4-6-8 powers for out-of-plane polarization, polarization gradient energy, electrostatic energy, elastic energy, electrostriction, surface energy, and flexoelectric contributions, is listed in **Appendix A2** [36].

Due to the incomplete screening, the presence of the dielectric gap can lead to the domain formation or a transition to the PE phase in the CIPS nanoflake. The dependence of the domain morphology and dynamics on the surface charge parameters, dielectric permittivity, mismatch strain, and sizes $R$, $h$, and $d$ can be studied numerically. The range of $d$, $h$, and $R$ values and surface charge parameters for which the single-domain state is a ground state in the CIPS nanoflake can be determined analytically and verified numerically.

Surface charges $\sigma_p$ and $\sigma_n$ partially screen the CIPS polarization in a complex nonlinear way (due to their density dependence on the surface electric potential given by Eqs.(3)). Since the appearance of the out-of-plane ferroelectric domains depends on the screening degree, the inhomogeneous distribution of the surface charges should be calculated numerically in a self-consistent way. If the domain structure is absent, the electric field does not depend on the transverse coordinates x and y, and therefore, a driving force leading to the inhomogeneous distribution of the surface charges is absent too.

In this case, the homogeneously distributed surface charges $\sigma_p$ and $\sigma_n$ partially screen the CIPS average polarization $\bar{P}$. Analytical results, presented below, are valid for the single-domain states (i.e., the FI and FII states) and the PE phase. The multi-domain states will be studied elsewhere.

In a single domain state and in the PE phase, the Landau-Khalatnikov equation determining the average polarization has the form [36]:

$$\Gamma \frac{d}{dt}\bar{P} + \tilde{\alpha}(T)\bar{P} + \tilde{\beta}\bar{P}^3 + \tilde{\gamma}\bar{P}^5 + \tilde{\delta}\bar{P}^7 = -\frac{\Psi}{h}. \tag{1a}$$

Here $\Gamma$ is the Khalatnikov coefficient, $\bar{P} = \frac{1}{h}\int_0^h P_3(\tilde{z})d\tilde{z}$ is the polarization over the nanoflake thickness, $\tilde{\alpha}$, $\tilde{\beta}$, $\tilde{\gamma}$, and $\tilde{\delta}$ are Landau expansion coefficients, and $\Psi$ is the overpotential. The Landau expansion coefficients are:

$$\tilde{\alpha}(T) = \alpha_T(T - T_C) + \frac{2g}{R\lambda + R^2/4} - 2\frac{Q_{13}(s_{22}-s_{12}) + Q_{23}(s_{11}-s_{12})}{s_{11}s_{22}-s_{12}^2}u_m, \tag{1b}$$

$$\tilde{\beta} = \beta + 2\frac{Q_{23}^2 s_{11} - 2Q_{13}Q_{23}s_{12} + Q_{13}^2 s_{22}}{s_{11}s_{22}-s_{12}^2} - 4\frac{s_{22}Z_{133} + s_{11}Z_{233} - s_{12}(Z_{133}+Z_{233})}{s_{11}s_{22}-s_{12}^2}u_m, \tag{1c}$$

$$\tilde{\gamma} = \gamma + 6\frac{Q_{13}s_{22}Z_{133} + Q_{23}s_{11}Z_{233} - Q_{13}s_{12}Z_{233} - Q_{23}s_{12}Z_{133}}{s_{11}s_{22}-s_{12}^2}, \tag{1d}$$

$$\tilde{\delta} = \delta + 4\frac{s_{22}Z_{133}^2 - 2s_{12}Z_{133}Z_{233} + s_{11}Z_{233}^2}{(s_{11}s_{22}-s_{12}^2)}. \tag{1e}$$

In Eq.(1a) $T$ is the temperature and $T_C$ is the Curie temperature of a bulk CIPS. The term $\frac{2g}{R\lambda + R^2/4}$ originates from the dipole-dipole correlation effect; $g$ is the positive polarization gradient coefficient, and $\lambda$ is the positive extrapolation length [43]. The terms in Eqs.(1) that are proportional to the mismatch strain $u_m$ and/or the combination of second order electrostriction coefficients $Q_{ij}$, higher order electrostriction coefficients $Z_{ijk}$, and elastic compliances $s_{ij}$, originate from the elastic strains coupling with the electrostriction [44, 45]. The values of $T_C$, $\alpha_T$, β, γ, δ, $Q_{ij}$, $Z_{ijkl}$ $s_{ij}$, $g$, and $\lambda$ are listed in **Table AI** in **Appendix A2** [36]. The values of $T_C$, $\alpha_T$, β, γ, δ, $Q_{ijkl}$, and $Z_{ijkl}$ were defined in Refs. [18, 19] from the fitting of the experimentally observed temperature dependence of dielectric permittivity [46, 47, 48], spontaneous polarization [14], and lattice constants [49] for hydrostatic and uniaxial pressures. The elastic compliances $s_{ij}$ were estimated from ultrasound velocity measurements [50, 51, 52].

Using the results of Ref. [28] and **Appendix A3** [36], the overpotential $\Psi$ can be determined in a self-consistent manner:

$$\frac{\Psi}{h} = \frac{d}{d\varepsilon_b + h\varepsilon_d}\frac{\bar{P}+\sigma}{\varepsilon_0} + \frac{\varepsilon_d U(t)}{d\varepsilon_b + h\varepsilon_d}. \tag{2a}$$

The total ionic-electronic charge $\sigma$ is the sum of positive ($\sigma_p$) and negative ($\sigma_n$) charges:

$$\sigma(\phi) = \sigma_p(\phi) + \sigma_n(\phi). \tag{2b}$$

From Eq.(2a), $\Psi$ contains a contribution which is proportional to the surface charge density $\sigma$, a depolarization field contribution which is proportional to $\bar{P}$, and an external potential drop which is proportional to the applied voltage $U(t)$ and inversely proportional to the CIPS thickness $h$. Below we consider the case of a periodic applied voltage, $U(t) = U \cdot \sin(\omega t)$, where $U$ is the amplitude, $\omega = \frac{2\pi}{\tau_V}$ is the frequency, and $\tau_V$ is the period of applied voltage.

According to SH approach, the nonlinear dynamics of the positive and negative surface charges obeys the relaxation equations [53]:

$$\tau_p \frac{\partial \sigma_p}{\partial t} + \sigma_p = eZ_p C_p \left(1 + g_p \exp\left(\frac{\Delta G_p^0 + eZ_p \Psi}{k_B T}\right)\right)^{-1}, \quad (3a)$$

$$\tau_n \frac{\partial \sigma_n}{\partial t} + \sigma_n = eZ_n C_n \left(1 + g_n \exp\left(\frac{\Delta G_n^0 + eZ_n \Psi}{k_B T}\right)\right)^{-1}. \quad (3b)$$

Here $e$ is an elementary charge; $Z_p$ and $Z_n$ are the ionization degrees of the positive and negative surface charges, respectively; and $C_p$ and $C_n$ are their 2D surface charge concentrations (measured in m$^{-2}$). Small $C_p$ and/or $C_n$ values correspond to the defect-free surface cleaned in the ultra-high vacuum. Large $C_p$ and/or $C_n$ can emerge because of defects, contaminations, oxygen excess or deficiency; they are defined by the adsorption isotherm in a more general case [24, 25]. Positive parameters $\Delta G_p^0$ and $\Delta G_n^0$ are the free energies of the surface defects formation under normal conditions and zero potential ($\phi = 0$). As a rule, the values of $\Delta G_{p,n}^0$ are poorly known and regarded as varying over the range (0.01 – 0.3) eV [24]. Positive prefactors $g_p$ and $g_n$ can originate from different mechanisms of the charge formation [54, 55].

Below we analyze the quasi-static solutions of nonlinear coupled equations, Eqs.(1)-(3), with equal concentrations of surface charges ($C_p = C_n$), equal prefactors ($g_p = g_n$), equal defect formation energies ($\Delta G_p^0 = \Delta G_n^0$), and opposite signs of the surface charges ($Z_p = -Z_n$). We re-designate these quantities as:

$$C_p = C_n = C, \quad Z_p = -Z_n = Z, \quad g_p = g_n = g, \quad \Delta G_p^0 = \Delta G_n^0 = \Delta G. \quad (3c)$$

The condition (3c) corresponds to the pairwise formation of negative and positive surface charges and agrees with the electroneutrality condition at $\phi = 0$ (see Eqs.(A.1)-(A.2) in **Appendix A1** [36] for details).

Also, we consider the simplest case of the same relaxation times, $\tau_p = \tau_n = \tau$, in Eq.(3a) and 3(b). In a general case, the relaxation times can be very different from each other for the case of ionic-electronic screening [31]. Since a polarization relaxation is determined by soft optical phonons, the strong inequality, $\Gamma/|\alpha| \ll \tau$, is valid far from the Curie temperature; and it makes sense to normalize time $t$ in Eqs.(3a)-(3b) to the Khalatnikov time, $\tau_{Kh} = \Gamma/|\alpha_T T_c|$. This normalization is used hereinafter.

Let us underline that the surface density of free charge on upper ($q_1$) and lower ($q_2$) electrodes can be measured experimentally. In **Appendix A3**, [36] we derived the following expressions for $q_1$ and $q_2$:

$$q_1 = -\frac{h\varepsilon_d}{d\varepsilon_b + h\varepsilon_d}\sigma - \frac{h\varepsilon_d}{d\varepsilon_b + h\varepsilon_d}\bar{P} + \frac{\varepsilon_0\varepsilon_b\varepsilon_d}{d\varepsilon_b + h\varepsilon_d}U, \tag{4a}$$

$$q_2 = -\frac{d\varepsilon_b}{d\varepsilon_b + h\varepsilon_d}\sigma + \frac{h\varepsilon_d}{d\varepsilon_b + h\varepsilon_d}\bar{P} - \frac{\varepsilon_0\varepsilon_b\varepsilon_d}{d\varepsilon_b + h\varepsilon_d}U. \tag{4b}$$

The sum $q_2 + q_1 + \sigma$ is always equal to zero, as it follows from the system electroneutrality condition. The total density of the surface free charge stored by the electrodes in the layered capacitor structure "CIPS nanoflake - surface charge layer - dielectric layer" is equal to $q = \frac{q_2 - q_1}{2}$. The charge density $q$, which can be measured under the total discharge of the capacitor structure charged by the voltage $U$, is equal to:

$$q(U) = \frac{h\varepsilon_d - d\varepsilon_b}{2(d\varepsilon_b + h\varepsilon_d)}\sigma + \frac{h\varepsilon_d}{d\varepsilon_b + h\varepsilon_d}\bar{P} - \frac{\varepsilon_0\varepsilon_b\varepsilon_d}{d\varepsilon_b + h\varepsilon_d}U. \tag{5}$$

The surface density of the stored free charge $q$ (measured in Coulombs per unit area) is an important characteristic for capacitors and energy storage devices. Below we name $q$ as the "stored charge" for the sake of brevity.

### III. RESULTS AND DISCUSSION
#### A. Ferri-ionic coupling in CIPS nanoflakes

In this work, we studied the numerical solution of Eqs.(1)-(3) for the voltage dependences of electric polarization, screening charges, and stored charge over a range of CIPS nanoflake sizes ($h\sim(2-100)$ nm, $R\sim(0.2-2)$ μm), dielectric layer thicknesses $d\sim(0-2)$ nm, concentrations of screening charges ($10^{-3}$nm$^{-2} \leq C \leq 10$ nm$^{-2}$), small formation energies of the screening charge, $\Delta G\sim(0.01-0.1)$ eV, and mismatch strains ($-1\% \leq u_m \leq +1\%$) in the vicinity of room temperature (285 K $< T <$ 300 K). The frequency $\omega$ of the applied voltage $U(t)$ was very low, $\omega\tau_{Kh} \ll 1$.

It was found that the behavior of the polarization is very sensitive to the mismatch strain $u_m$, nanoflake thickness $h$, and charge concentration $C$; it is less sensitive to the formation energy $\Delta G$. It appeared that the most interesting results correspond to the strain range $-0.3\% \leq u_m \leq +0.6\%$, which is the same range of the phase diagram of the CIPS film covered by electrodes (see e.g., Figs. 2 and 3 in Ref. [56]). The traces outside the strain range $-0.3\% \leq u_m \leq +0.6\%$ do not show any qualitatively different behavior and therefore are not shown in **Figs. 2 – 5**. Since the Curie temperature of bulk CIPS is 292.7 K, the most interesting observations correspond to the vicinity of room temperature.

Typical quasi-static hysteresis loops of the average polarization $\bar{P}(U)$, positive and negative surface charges, $\sigma_p(U)$ and $\sigma_n(U)$, and the stored charge $q(U)$ are shown in **Fig. 2** and **Fig. 3**, for $h = $ 10 nm and 100 nm, respectively. The dependences of remanent polarization $\bar{P}(0)$ and stored charge $q(0)$ on $C$ and $u_m$ are shown in **Fig. 4** for $h$ =10 nm and 100 nm. The quasi-static hysteresis loops plotted for a large number of $C$ and $u_m$ values are presented in **Fig. A2** and **A3**, and the dependences of maximal polarization and stored charge on $C$ and $u_m$ are presented in **Fig. A4** in **Appendix A4** of the Supplemental Material [36].

Hysteresis loops for 10-nm nanoflakes covered with surface charges in smaller concentrations $C \leq 0.01$ nm$^{-2}$, are shown in **Fig. 2** (left side). A change of the mismatch strain from compressive strain (top row, $u_m = -0.25\%$) to tensile strain (bottom row, $u_m = +0.6\%$) leads to the following gradual transformation of $\bar{P}(U)$. A rectangular-shaped single hysteresis loop ($u_m = -0.25\%$) transforms into a double loop ($u_m = -0.1\%$), and then into a hysteresis-less paraelectric curve ($u_m = 0\%$) that persists until larger tensile strains ($u_m = +0.6\%$) are applied. Due to the definite signs of the electrostriction coefficients in CIPS (see **Table AI**), compressive strains support its spontaneous polarization $P_3$ better than tensile strains, and because of this the polarization recovering to a single loop at larger tensile strains ($u_m \geq 0.25\%$) is not as pronounced as for the same compressive strain ($u_m \leq -0.25\%$).

Hysteresis loops for 10-nm nanoflakes covered with surface charges in higher concentrations $C \geq 0.1$ nm$^{-2}$, are shown in **Fig. 2** (right side). A change of $u_m$ from $-0.25\%$ (top row) to $+0.6\%$ (bottom row) leads to the following gradual transformation of $\bar{P}(U)$. A wide rectangular-shaped single hysteresis loop becomes narrower, and then transforms into a slightly pinched loop, a strongly pinched loop, a double loop, a hysteresis-less paraelectric curve. The hysteresis-less curves transform back to narrow single hysteresis loops with relatively weaker polarizations; the thin loops become wider and acquire a pronounced rectangular-like shape with a further increase of $u_m$ above 0.5 %, which is shown in the bottom row.

Unlike the case of smaller concentrations ($C \leq 0.01$ nm$^{-2}$), the stored charge loops differ from the polarization loops for the higher concentrations investigated ($C \geq 0.1$ nm$^{-2}$): for the compressive strains ($u_m < 0$) and high tensile strains ($u_m \geq +0.6\%$) the $q$ magnitude reaches (2 – 12) µC/cm² with a voltage increase, which is significantly higher than the saturated polarization with values less than (1 – 4) µC/cm² (compare the black and green loops in **Fig. 2** (right side)). Zero-field and saturation values of $\sigma_p(U)$ and $\sigma_n(U)$ are relatively high and strongly increase with an increase in $C$. Simultaneously with the increase of $\sigma_p(U)$ and $\sigma_n(U)$, the difference between the shape of $\bar{P}(U)$ and $q(U)$ hysteresis loops increases.

The coercive voltage for $\bar{P}$ and $q$ strongly increases (up to several times) with an increase in $C$ from 0.001 nm$^{-2}$ to 1 nm$^{-2}$ for the strongest compressive strains (see the top row in **Fig. 2**). For the strongest tensile strains, the coercive voltage increases more than 10 times with the same increase in $C$ (see the bottom row in **Fig. 2**). Hereinafter we introduce the "electric coercive voltage" of the physical quantity "X", which reveals a pronounced hysteresis behavior in the applied electric voltage $U$. The coercive voltage is defined as the voltage $U_c$ for which $X(U_c)$ changes its sign (i.e., $X(U_c) = 0$), and therefore, $U_c$ can exist for $\bar{P}$ and $q$ (which can change their sign) and is not applicable for the positive and negative charges (by definition). Let us underline that the domain formation, which is not considered in this work, can significantly decrease the coercive field (and thus the coercive voltage). As a rule, an observed coercive field is defined by the domain walls motion and pinning effects and can be much smaller than the thermodynamic coercive field calculated for the single-domain polarization switching.

It is worth noting that the weak (for small $C$) or moderate (for large $C$) changes in double loop width, height, and area, which occur in unstrained CIPS nanoflakes with an increase in $C$, are the pure manifestation of the coupling between the ferrielectric dipoles and surface ionic charge (see the row for $u_m = 0$ in **Fig. 2**). The significant changes of the loop shape, width, height, and area, which occur in strained CIPS nanoflakes with the increase in $C$ are the joint action of the strain (as a main factor) and the ferro-ionic coupling. Furthermore, the high values of the maximal stored charge, $q \sim (10 - 14)$ µC/cm$^2$ for $C \cong 1$ nm$^{-2}$ for the case of compressive strains, can mimic the field-induced mid-gap polarization states in CIPS, however, the high stored charge is merely high-field ferri-ionic states in thin CIPS nanoflakes.

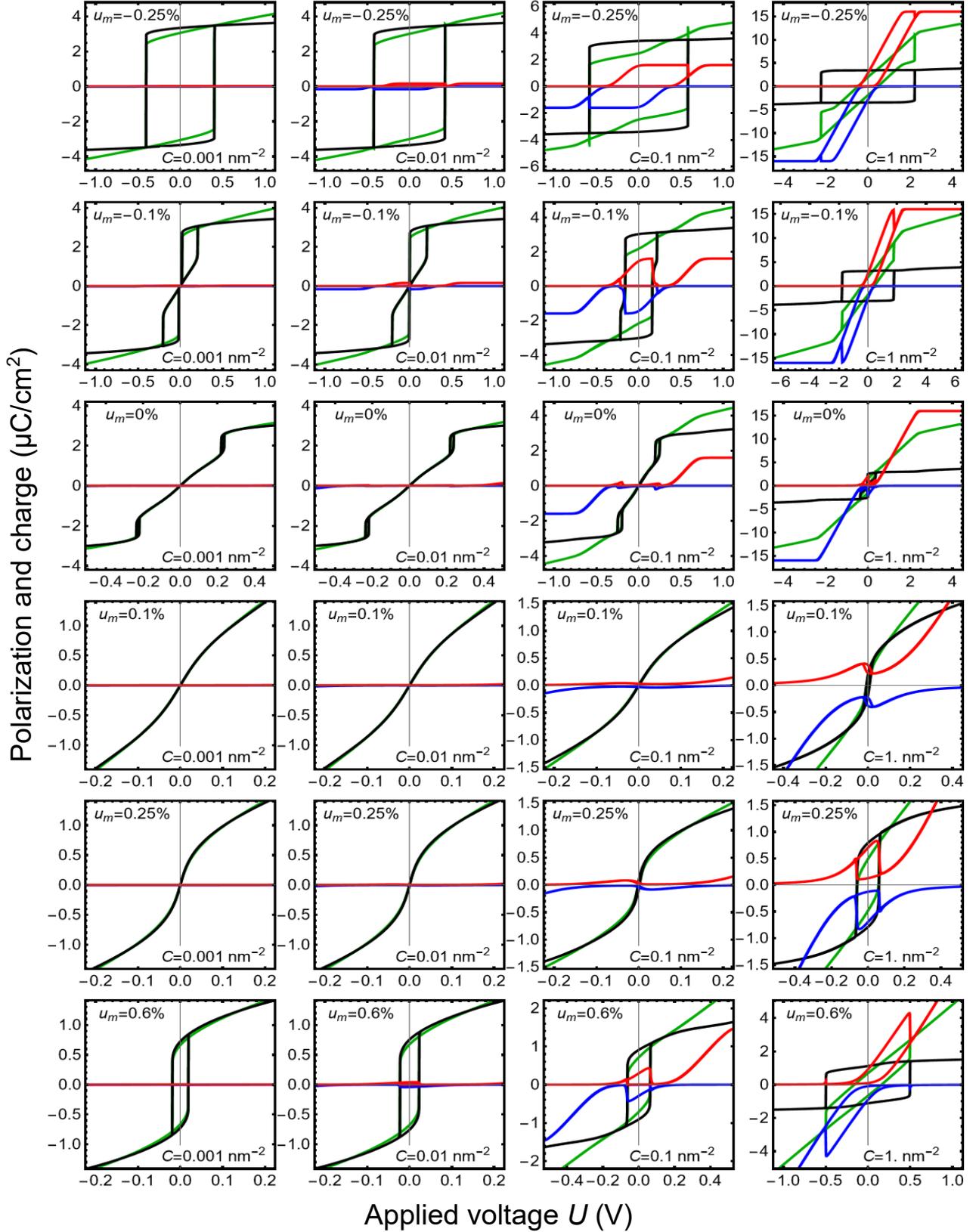

**FIGURE 2.** The quasi-static voltage dependences of the average polarization $\bar{P}(U)$ (black curves), positive (red curves) and negative (blue curves) surface charges, $\sigma_p(U)$ and $\sigma_n(U)$, and stored charge $q(U)$ (green curves),

calculated for 10-nm CIPS nanoflakes with different concentrations of surface charges $C$, varying from $10^{-3}$ nm$^{-2}$ to 1 nm$^{-2}$, and different degrees of mismatch strain $u_m$, ranging from -0.25% to +0.6%. Corresponding values of $C$ and $u_m$ are listed inside each plot. Other parameters: $h = 10$ nm, $R \geq 0.5$ μm, $d =1$ nm, $\varepsilon_d = 10$, $\varepsilon_b = 9$, $g = 1$, $\Delta G = 0.1$ eV, $\tau_n = \tau_p = 10\tau_{Kh}$, and $T = 293$ K.

Hysteresis loops for thicker nanoflakes (100 nm) covered with surface charges are shown in **Fig. 3**. A change of mismatch strain from compressive strain (top row, $u_m = -0.25$ %) to tensile strain (bottom row, $u_m = +0.25$ %) leads to the following gradual transformation of $\bar{P}(U)$. A single rectangular-shaped hysteresis loop transforms into a pinched loop, then as the pinch tightens a strongly pinched loop transforms into a triple loop. The triple loop looks like a double loop, because its central loop is very thin and hardly visible in comparison with the two pronounced side loops (see $u_m = 0.05\%$ for smaller surface charge concentrations). The double-like triple loop transforms back to a single loop with a further increase of $u_m$ ($u_m \geq 0.1\%$). The single loops become wider and acquire a pronounced rectangular-like shape with a further increase of $u_m$ above 0.25 %, which is not shown in **Fig. 3.**

For small concentrations of surface charge ($C \leq 0.01$ nm$^{-2}$) the stored charge loops are similar to the polarization loops; and a difference between them increases with increasing values of $C$ (compare the black and green loops on the left side of **Fig. 3**). The values of $\sigma_p(U)$ and $\sigma_n(U)$ are relatively small and increase with an increase in $C$. For higher concentrations, $C \geq 0.1$ nm$^{-2}$, the zero-field and saturation values of $\sigma_p(U)$ and $\sigma_n(U)$ are relatively high and strongly increase with an increase in $C$ (compare the red and blue loops on the right side of **Fig. 3**). The stored charge loops differ significantly from the polarization loops: the magnitude of $q(U)$ reaches $(4 - 12)$ μC/cm$^2$, being much higher than the polarization in saturation, which is less than $(2 - 4)$ μC/cm$^2$ (compare the black and green loops on the right side of **Fig. 3**). The maximal values of the stored charge can be very high, $q(U) \cong (10 - 15)$ μC/cm$^2$ for $C \cong 1$ nm$^{-2}$, as shown in the right column of **Fig. 3**.

Note, that the 10 times increase in the coercive voltage $U_c$ of the single loops for the 100-nm nanoflakes in comparison with the 10-nm nanoflakes (compare, e.g., the first columns in **Fig. 3** and **Fig. 2**) corresponds to the same coercive field $E_c$, because the proportionality law $E_c \cong \frac{U_c}{h}$ is valid for the single-domain polarization switching. However, the domain formation in weakly screened nanoflakes (i.e., for a small concentration $C$ of surface charges) can easily break the proportionality law.

## Hysteresis loops for the 100-nm CIPS nanoflake

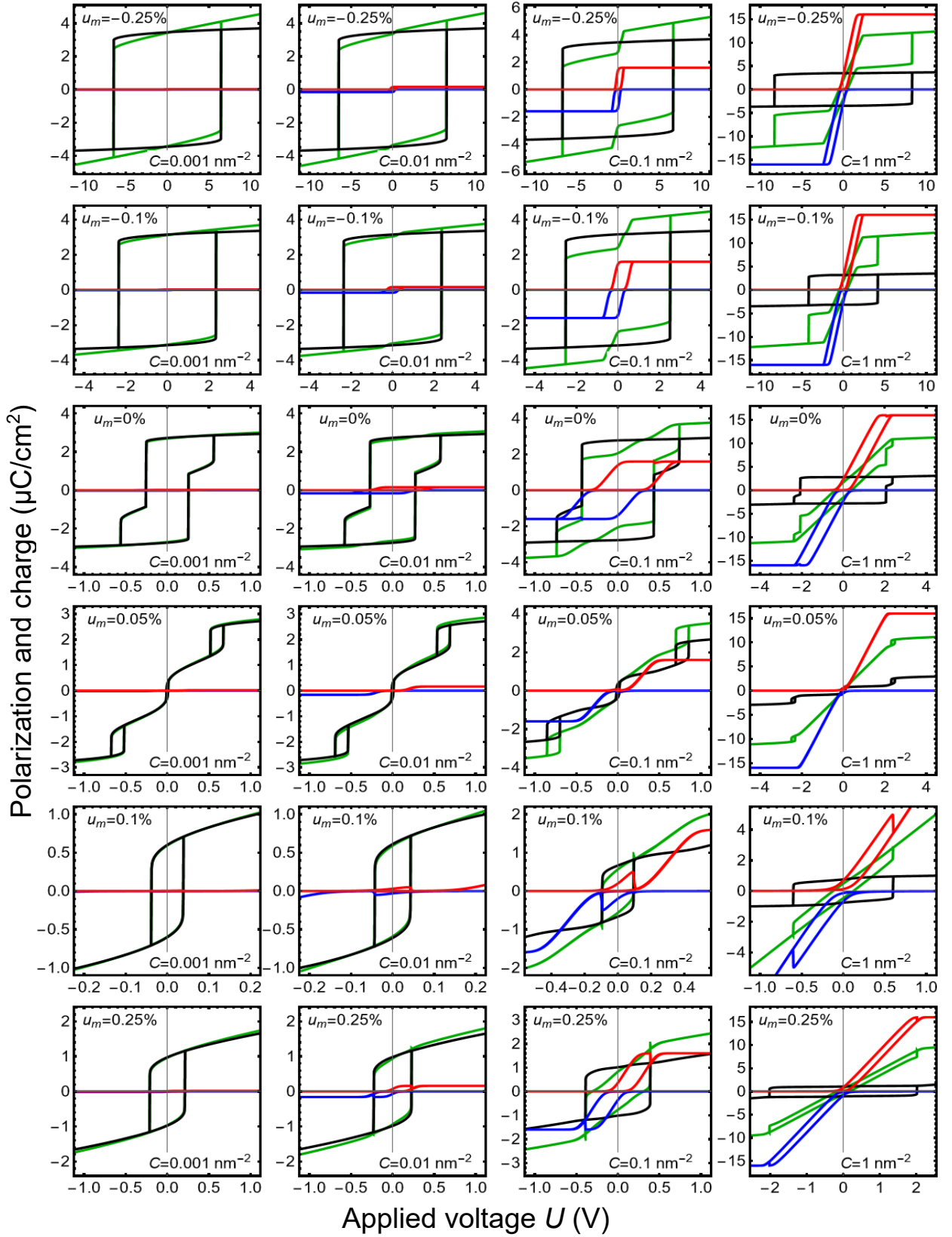

**FIGURE 3.** The quasi-static voltage dependences of the average polarization $\bar{P}(U)$ (black curves), positive (red curves) and negative (blue curves) surface charges, $\sigma_p(U)$ and $\sigma_n(U)$, and stored charge $q(U)$ (green curves),

calculated for 100-nm CIPS nanoflakes with different concentrations of surface charges $C$ varying from $10^{-3}$ nm$^{-2}$ to 1 nm$^{-2}$, and different degrees of mismatch strain $u_m$, ranging from -0.25% to +0.25%. Corresponding values of $C$ and $u_m$ are listed inside each plot. $h = 100$ nm and $R \geq 5$ μm. Other parameters are the same as in **Fig 2**.

One can conclude that the polarization of thinner nanoflakes reveals a much stronger dependence on the surface charge concentration and mismatch strain, particularly for the tensile strains. The physical origin of this dependence is due to the larger surface to volume ratio in comparison to thick flakes. The ratio is proportional to the flake thickness $h$. According to Eq.(2a), the size effect of acting electric field scales as $\frac{d}{h}$ for $h \gg d$, and thus we can expect a $1/h$ scaling law in the surface charge impact on the flake properties.

The key differences between **Fig. 3** and **Fig. 2** are the existence of double-like triple loops and the apparent absence of hysteresis-less curves for 100-nm CIPS nanoflakes in comparison with 10-nm CIPS nanoflakes near the room temperature (at 293 K). The differences that exist near room temperature are related to the very small region of the PE phase in 100-nm nanoflakes in comparison with the much larger region of the PE phase in 10-nm nanoflakes. Indeed, the formation of double loops only occurs when transforming directly out of the PE phase, and, since the phase is nearly absent near the room temperature in the relatively well-screened large 100-nm nanoflakes, the double loops are almost absent. Namely, additional calculations for 100-nm nanoflakes show that double-like loops exist in the very narrow strain range between 0.04% and 0.045% (not shown in **Fig. 3**). Instead of the double loops, the double-like triple loops and/or strongly pinched loops can be formed in the low-polarized ferrielectric state of 100-nm CIPS nanoflakes.

It is also worth noting that the manifestations of ferri-ionic coupling are much stronger for 10-nm nanoflakes, where the ferri-ionic states exist over a wider range of $C$ and $u_m$ values. This behavior demonstrates the significant roles that both surface charge concentration and strain play on the phase, and more particular, how it affects the ability to achieve different types of hysteresis loops that are further defined in the phase diagrams in **Fig. 4**, which are described below.

**Figures 4(a)** and **4(b)** show the significant influence of mismatch strain $u_m$ and surface charge concentration $C$ on the remanent polarization $\bar{P}(0)$ and the stored charge $q(0)$ for the 10-nm CIPS nanoflake at 293 K and $U = 0$. For small concentration values ($C < 0.01$ nm$^{-2}$), compressive strains ($u_m < -0.1\%$) induce and support the FI1 ferrielectric state with a relatively large spontaneous polarization (>3.0 μC/cm$^2$) and stored free charge (>2.5 μC/cm$^2$); meanwhile tensile strains ($u_m >$ 0.35 %) induce and support the FI2 ferrielectric state with a relatively small spontaneous polarization (<1.0 μC/cm$^2$) and stored free charge (<0.5 μC/cm$^2$). For very small concentrations of surface charges ($C < 0.001$ nm$^{-2}$) the PE phase, where $\bar{P}(0) = 0$, is stable in the range of mismatch strains, $-0.1\% < u_m < 0.37\%$. The transition from the FI1 state to the PE phase is of the first order, and the transition from the PE phase to the FI2 state is of the second order. The area of the PE phase region gradually decreases with an increase in $C$ and nearly disappears for $C > 5$ nm$^{-2}$. Simultaneously with the vanishing of the PE phase, the FI1 and FI2 states gradually transform into the ferri-ionic state 1 (FII1) with higher stored charge (~2 μC/cm$^2$) and the ferri-ionic state 2 (FII2) with lower stored charge (~1 μC/cm$^2$). At $C > 5$ nm$^{-2}$ the boundary between the FII1 and FII2 states is sharp indicating on the first order phase transition.

It is worth noting that the labeling for FI and FII states are only present in **Fig. 4(a)** for polarization (as well as in **Fig. 4(c)**). The labels mark the states of electric polarization related with the existence of two dipole sublattices (as explained in the Introduction), which can be weakly (for small $C$) or strongly (for higher $C$) coupled to the ionic-electronic charge density $\sigma$ accumulated at the CIPS surface. The polarization coupled with the surface charges is the physical state of dipoles and surface ions, which defines the magnitude of the total charge density $q$ stored in the layered capacitor. In fact, $q$ is a working performance of the capacitor, not a physical state; therefore, there is no classification for it in terms of polarization states.

**Figures 4(c)** and **4(d)** show a strong influence of mismatch strain on the remanent polarization $\bar{P}(0)$ and the stored charge $q(0)$ for the 100-nm CIPS nanoflake at 293 K. The main differences of these figures from **Figs. 4(a)** and **4(b)** are the ultra-thin region of the PE phase for the 100-nm nanoflakes, and the very weak dependence of the remanent polarization, $\bar{P}(0)$, on the surface charge concentration $C$. This is because the influence of the depolarization field $E_d$ and surface charge $\sigma$ on $\bar{P}(0)$ and $q(0)$ are proportional to the thickness-dependent factor $\frac{d}{h}$. Indeed, the depolarization effect is described by Eq.(2a), where the depolarization field $E_d = -\frac{d}{d\varepsilon_b + h\varepsilon_d}\frac{\bar{P}+\sigma}{\varepsilon_0}$ is included in the right-hand side of equation. Since $h \gg d$, we obtain $E_d \sim \frac{d}{h}(\bar{P} + \sigma)$. In this case, $E_d$ is inversely proportional to the nanoflake thickness $h$, and directly proportional to the concentration $C$, since $\sigma \sim C$ in accordance with Eqs.(3). Hence, due to the $1/h$ proportionality, depolarization and ionic screening effects are much stronger for 10-nm nanoflakes compared to 100-nm nanoflakes.

Due to the weak dependence on $C$, the FI1 and FI2 states, as well as the FII1 and FII2 states, are separated with a relatively sharp boundary in **Figs. 4(c)** and **4(d)**, which is nearly a vertical line at $u_m \approx 0.03\%$. All other features shown in **Figs. 4(c)** and **4(d)** are similar to those shown in **Figs. 4(a)** and **4(b)**.

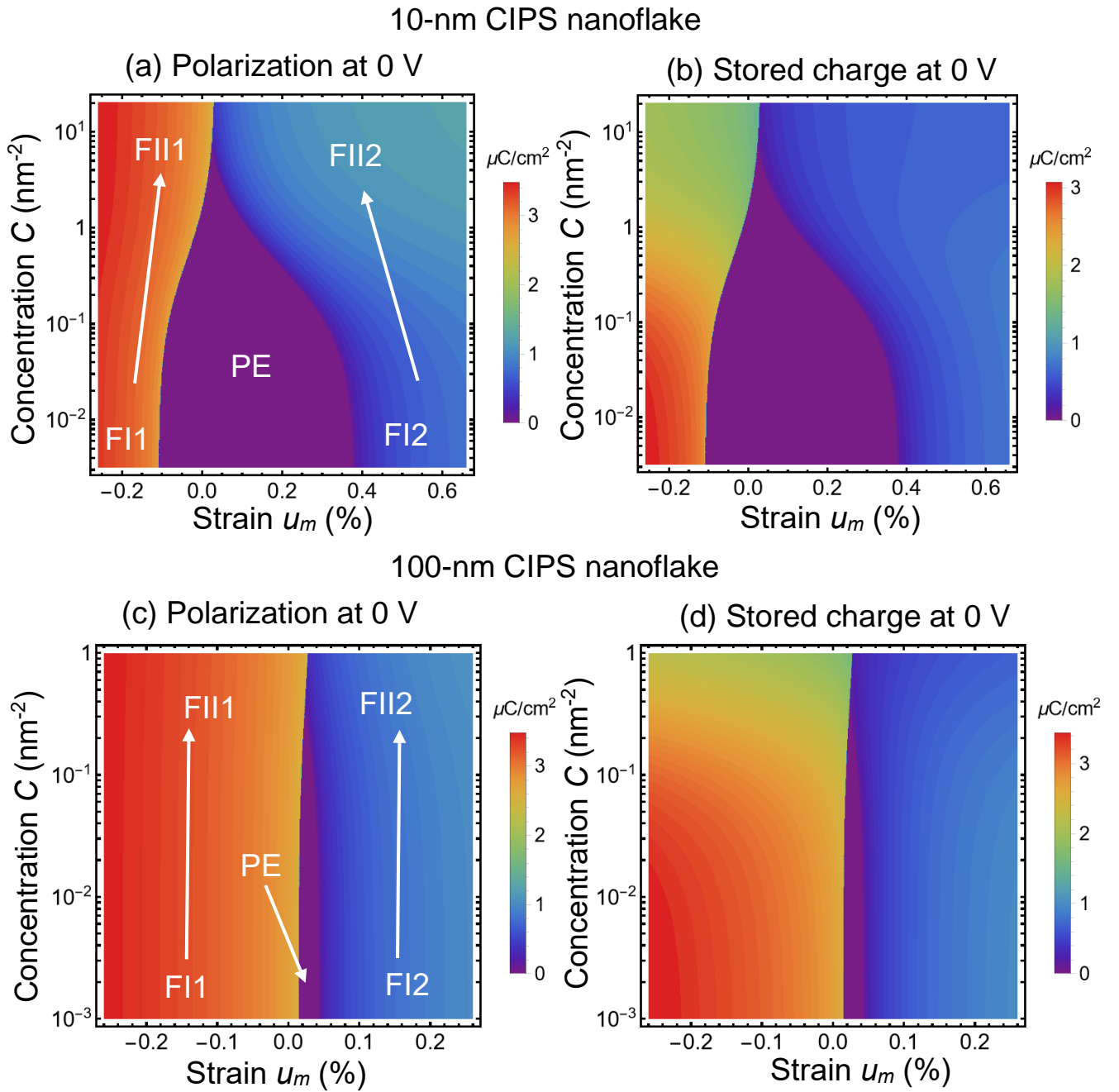

**FIGURE 4.** The dependences of the remanent polarization $\bar{P}(0)$ **(a, c)** and stored charge $q(0)$ **(b, d)** on concentration of surface charges $C$ and mismatch strain $u_m$ for 10-nm **(a, b)** and 100-nm **(c, d)** CIPS nanoflakes. Other parameters are the same as in **Fig 2**. Abbreviations "PE", "FI1", "FI2", "FII1", and "FII2" represent the paraelectric phase, the ferrielectric state 1 (with higher $\bar{P}$ and small $C$), the ferrielectric state 2 (with lower $\bar{P}$ and small $C$), the ferri-ionic state 1 (with higher $\bar{P}$ and large $C$), and the ferri-ionic state 2 (with lower $\bar{P}$ and large $C$), respectively.

Results shown in **Fig. 4** and **Fig. A4** [36] correlate with the different hysteresis loops presented in **Figs. 2** and **3**. They provide additional insight to the interpretation of the hysteresis loops. It is clearly seen that different shaped loops are the signature not only for different phases, but also for significantly different polarization strengths, even when considering the same phase, e.g., FI1 and FI2. While the FI1 and FII1 states with high $\bar{P}$ values are required for the charge storage, the FI2 and FII2 states with very low $\bar{P}$ (near the flat wells), as well as the PE phase that is enhanced via tensile mismatch strain in CIPS nanoflakes, can reveal a negative differential capacitance effect like in many ferroelectric films. In the next section we show that the NC effect appears to be much greater in CIPS compared to other common ferroelectric materials.

## B. The negative capacitance effect in CIPS nanoflakes

The stabilization of ferroelectric films, such as lead zirconate-titanate ($Pb_xZr_{1-x}TiO_3$), in the state of negative capacitance (NC) [57] was revealed experimentally more than ten years ago [58]. In particular, Khan et. al. demonstrated that the total capacitance of a double-layer capacitor, made of paraelectric strontium titanate ($SrTiO_3$) and ferroelectric $Pb_xZr_{1-x}TiO_3$, is greater than it would be for a single-layer capacitor comprising $SrTiO_3$ of the same thickness as used in a double-layer capacitor. Using the ferroelectric NC insulator of an appropriate thickness in the gate stack of a Field-Effect Transistor (FET) has several advantages for its architecture, because the NC insulator requires less energy, which significantly reduces heating of nano-chips with a high density of critical electronic elements [59].

Many experimental demonstrations of the NC effect in ferroelectric double-layer capacitors are available [60, 61, 62, 63], but only a few propose semi-analytical expressions for the conditions of the NC effect appearance and consider the inevitable appearance of the domain structure in the ferroelectric layer (see e.g., Refs. [64, 65, 66, 67]). The disproportion between the experimental advances and their analytical description is explained as follows: it is very difficult to find the analytical conditions of the NC effect appearance and stability, which include the search of a proper ferroelectric material, its geometry, the working temperature, and thickness ranges.

In this regard, CIPS thin films and nanoflakes seem very promising candidates for the analytical control of the NC effect, which has been observed in FETs based on CIPS [68, 69, 70]. Indeed, Wang et al. [68] demonstrated the steep-slope NC-FETs with the channel made of two-dimensional $MoS_2$ and a CIPS thin film as a gate oxide. The NC-FET has an average subthreshold swing less than the Boltzmann's limit (60 mV/decade at room temperature) for over seven decades of drain current, and a negligible hysteresis of the subthreshold swing is achieved for a thickness of CIPS less than 20 nm. Dey et al. [69] demonstrated that the NC effect in CIPS can stabilize logic states in tunnel FETs.

The analytical control of the NC effect is easily possible in CIPS films, because their 8-th order thermodynamic potential can contain four wells, which flatten in the vicinity of the so-called "critical end point" (CEP) and "bicritical end point" (BEP) [9, 10, 18]. It is important that the temperature position of the CEP and BEP can be easily controlled by the flake-substrate lattice mismatch [9, 18]. Appearance of the "flat wells", which contain a great number of energy-degenerate states with a small nonzero polarization, can be the reason for the pronounced NC effect; this effect may appear when the flat wells split into a very wide plateau in the presence of a dielectric layer.

The goal of this subsection is to show that the presence of the dielectric layer and the ionic-electronic surface charge can induce a PE-like state with a plateau-like potential energy well located near zero energy values. The PE-like state is formed instead of the PE phase with flat-well states (e.g., FI2 and/or FII2) and a very low spontaneous polarization. The PE-like state, where the NC effect is pronounced, can exist over a specific range of dielectric layer and flake thicknesses, mismatch strains, and surface charge densities. Below we derive the analytical conditions for this specific range of parameters.

Indeed, our analytical calculations and FEM show that the spontaneous polarization of CIPS nanoflakes, whose surface is covered by an ionic-electronic charge layer, can be "stabilized" in the energy-degenerated metastable states, some examples of which (for a single-domain bulk material) are schematically shown by the green curve in **Fig. 5(a)**. The free energy potential of these states has relatively flat negative wells, which couple to the positive parabolic potential of the dielectric layer (DL) (see the blue curve in **Fig. 5(a)**). In the presence of ionic surface charge (SC), the total potential relief of the bi-layer system can become significantly flatter than the dielectric layer potential for certain layer thicknesses, mismatch strains, and surface charge densities (see the red curve in **Fig. 5(a)**). As a result, the free charge $q = \frac{q_2 - q_1}{2}$ stored at the electrodes covering the system "CIPS+SC+DL" becomes larger than the "reference" charge at the electrodes covering the dielectric layer. The effective differential capacitance $C_{eff}$ of the bi-layer capacitor of thickness $d + h$ (defined in **Fig. 1**) is equal to the first derivative of $q$ over the applied voltage $U$: $C_{eff} = \frac{dq}{dU}$. The NC effect occurs when the value of $C_{eff}$ becomes greater than the capacitance $C_r$ of the reference dielectric capacitor of thickness $d$. In **Appendix A5** [36] we derived the expression for the relative differential capacitance:

$$\frac{\Delta C}{C_r} = \frac{C_r - C_{eff}}{C_r} = \left(\frac{h\varepsilon_d - d\varepsilon_b}{d\varepsilon_b + h\varepsilon_d}\frac{d\sigma}{dU} + \frac{2h\varepsilon_d}{d\varepsilon_b + h\varepsilon_d}\frac{d\bar{P}}{dU} - 2\frac{\varepsilon_0 \varepsilon_b \varepsilon_d}{d\varepsilon_b + h\varepsilon_d} - \frac{\varepsilon_0 \varepsilon_d}{d}\right), \tag{6}$$

where we used the expression $C_r = \frac{\varepsilon_0 \varepsilon_d}{d}$.

The voltage dependence of the bi-layer capacitor relative differential capacitance, $\frac{\Delta C}{C_r}$, calculated for $C = 10^{-3}$ nm$^{-2}$, $h = 10$ nm, and mismatch strain $u_m = 0.3$ % is shown in **Fig. 5(b)**. The NC effect (when the capacitance change $\Delta C < 0$) exists in the small voltage regime; this regime permits energy saving nano-chips with high density of working elements for NC effect applications.

The dependence of the ratio $\frac{\Delta C}{C_r}$ on the concentration of surface charges $C$ and mismatch strain $u_m$ is shown in **Fig. 5(c)** and **Fig. 5(d)** for the 10-nm and 100-nm CIPS nanoflakes, respectively. It is seen from the figures that the tensiled 10-nm nanoflakes are much more promising for the realization of the NC effect.

For 10-nm CIPS nanoflakes, the NC effect exists in a relatively large right triangular-like region (marked by thin black curves in **Fig. 5(c)**), where the vertex is located near the point $C \approx 1$ nm$^{-2}$ and $u_m \approx 0.03\%$. The vertex point is very close to the point of the PE phase disappearance in **Fig. 4(a)**. The triangle base corresponds to the tensile mismatch strains $0 < u_m < 0.55\%$ and small concentrations $C \leq 10^{-3}$ nm$^{-2}$. The increase of $C$ leads to the shrinking of the NC region up to its complete disappearance for $C > 1$ nm$^{-2}$. The location and shape of the NC effect region in **Fig. 5(c)** correlate with the location and shape of the size-induced PE phase (with zero spontaneous polarization) and the FII2 state (with a small polarization less than 0.1 µC/cm$^2$) in **Fig. 4(a)**.

For 100-nm flakes, the NC effect exists in a very thin needle-like region (marked by thin black curves in **Fig. 5(d)**), where the vertex is located near the point $C \approx 0.04$ nm$^{-2}$ and $u_m \approx 0.02\%$. The maximal width of the NC region corresponds to the tensile mismatch strains $0.02\% < u_m < 0.06\%$. The very thin region of the NC effect in **Fig. 5(d)** occupies part of the very thin PE region ($|\bar{P}(0)| = 0$) and part of the FII2 region with a small spontaneous polarization ($|\bar{P}(0)| < 0.1$ µC/cm$^2$) in **Fig. 4(c)**.

The difference in the NC effect behavior for 10-nm and 100-nm CIPS nanoflakes is explained by the fact that NC can exist only in the flakes with the very small and zero spontaneous polarization, $|\bar{P}(0)| < 0.1$ µC/cm$^2$. In accordance with our calculations, the regions with small and zero $\bar{P}(0)$ are big enough in (5 – 20) nm-thick CIPS nanoflakes and become very small in (50 – 100) nm-thick CIPS nanoflakes. These results are in a qualitative agreement with experiments [68, 69], which also demonstrate the great potential of thin CIPS nanoflakes for flexible ultra-low-power NC FETs and tunnel junctions.

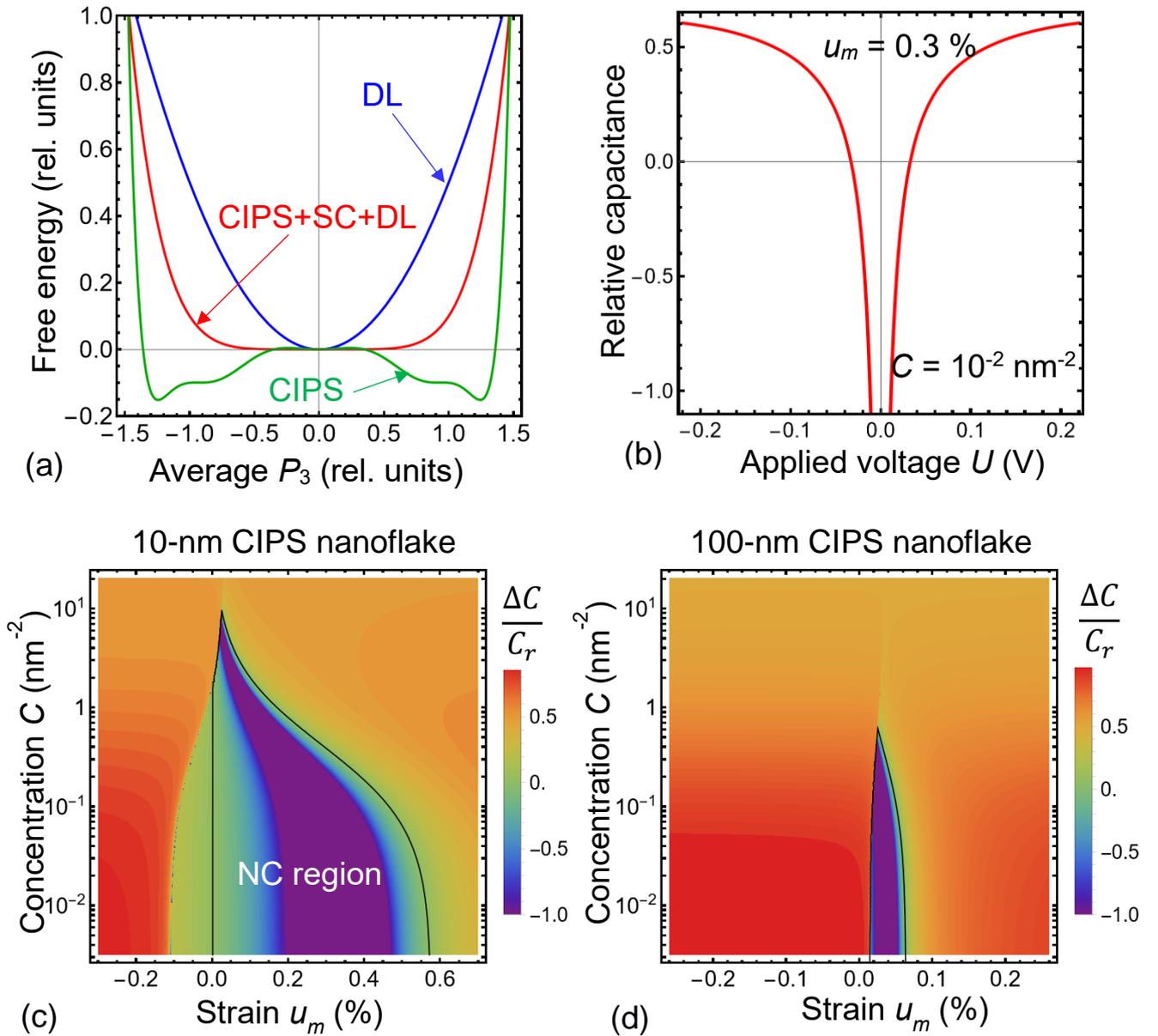

**FIGURE 5. (a)** Schematic illustration of the free energy dependence on the polarization for single-domain bulk CIPS with metastable polarization states (green curve marked as "CIPS"), the dielectric layer (blue curve marked as "DL"), and the bi-layer structure consisting of the CIPS nanoflake covered by surface charge and a dielectric layer (red curve marked as "DL+SC+CIPS"). **(b)** The voltage dependence of the bi-layer capacitor relative capacitance, $\frac{\Delta C}{C_r}$, calculated for $h = 10$ nm, $C = 10^{-2}$ nm$^{-2}$, and mismatch strain $u_m = 0.3$ %. **(c, d)** The dependence of the dimensionless ratio, $\frac{\Delta C}{C_r}$, on the concentration of surface charge $C$ and mismatch strain $u_m$. The CIPS nanoflake sizes: $h = 10$ nm and $R \geq 0.5$ μm for plot **(c)**; $h = 100$ nm and $R \geq 5$ μm for plot **(d)**. Other parameters are the same as in **Fig. 2**.

## IV. CONCLUSIONS

Nanoflakes of van der Waals ferrielectric CIPS covered by ionic-electronic charge layers can exist in the ferri-ionic polar state with a relatively high polarization ~ 5 $\mu$C/cm$^2$ and stored free charge ~ 10 $\mu$C/cm$^2$, which can mimic "mid-gap" states related to the surface field-induced transfer of Cu and/or In ions in the van der Waals gap. The physical origin of the ferri-ionic state is the nonlinear interaction between the ferroelectric dipoles and surface charges with slow relaxation dynamics in an external field. The interaction leads to the emergence of a broad range of paraelectric, paraelectric-like, ferrielectric, and ferri-ionic states in CIPS nanoflakes. While the ferrielectric and ferri-ionic states with high polarization values are required for charge storage, the states with very low polarization, as well as the PE phase that is enhanced via a tensile mismatch strain in CIPS nanoflakes, can reveal a more efficient NC effect compared to many popular ferroelectric films. The crossover between these states can be controlled by mismatch strain, size effects, and characteristics of surface charges in the applied field.

The states' stability and/or metastability are determined by the minimum of the system free energy consisting of the coupled screening charges energy, depolarization field energy, elastic energy, and the Landau-type four-well potential of the dipole polarization. The possibility to govern the transitions between the paraelectric, ferrielectric, and ferri-ionic states by mismatch strain and ionic-electronic screening can be useful for controlling the tunneling barrier in thin film devices based on CIPS nanoflakes. The proposed analytical description allows for the calculation of a range of parameters (e.g., dielectric layer thickness and permittivity, CIPS nanoflake thickness, mismatch strains, and characteristics of the surface charges) for which the switching between paraelectric-like states with a very small (or absent) spontaneous polarization and ferri-ionic states with a high polarization can be used in the resonant tunnel junctions and/or other tunneling-based thin film devices. In particular, compressive mismatch strains (less than -0.1 %) and high concentrations of surface charges (more than 0.1 nm$^{-2}$) support the stable high-polarization ferri-ionic state in (20 – 100) nm thick CIPS nanoflakes.

We predict that the CIPS nanoflakes reveal features of a controllable NC effect in the range of parameters, which are quite different from those required for the stability of high-polarization states (e.g., tensile mismatch strains, low concentrations of surface charges, and ultra-thin flakes). We predict this to be the case because the 8-th order thermodynamic potential of CIPS nanoflakes can contain four wells, which flatten in the vicinity of special (critical and bicritical) end points. Appearance of the "flat wells", which contain a great number of energy-degenerate states with a small nonzero polarization, is the reason for the pronounced NC effect, which appears when the flat wells split into a very wide plateau in the presence of a dielectric layer. The temperature position of the critical points can be controlled by the flake-substrate lattice mismatch.

It appears that thin (e.g., 5 – 20 nm) nanoflakes are much more promising for the realization of the NC effect in comparison with thick (e.g., 50 – 100 nm) nanoflakes. This theoretical prediction is in qualitative agreement with the experimental results [68, 69]. For thin nanoflakes, the NC effect exists at room temperature for a relatively wide range of tensile mismatch strains (from 0 % to 0.55 %) and small concentrations of surface charges (less than 0.3 nm$^{-2}$). The increase of the surface charge concentration leads to the shrinking of the triangular-like NC region up to a point of its complete disappearance. For thick nanoflakes, the NC effect exists at room temperature in a very thin needle-like region for tensile mismatch strains less than 0.05%. The difference in the NC effect behavior for thin and thick CIPS nanoflakes is explained by the fact that the NC can exist only in the regions of very small or zero spontaneous polarization. The controllable NC effect makes thin CIPS nanoflakes attractive for nano-capacitors and promising for gate oxide materials with better energy efficiency and reduced heat generation.

**Acknowledgements.** A.N.M. acknowledges EOARD project 9IOE063b and related STCU partner project P751b. This effort (problem statement and general analysis, S.V.K.) was supported as part of the center for 3D Ferroelectric Microelectronics (3DFeM), an Energy Frontier Research Center funded by the U.S. Department of Energy (DOE), Office of Science, Basic Energy Sciences under Award Number DE-SC0021118. The work of E.A.E. is supported by the DOE Software Project on "Computational Mesoscale Science and Open Software for Quantum Materials", under Award Number DE-SC0020145 as part of the Computational Materials Sciences Program of US Department of Energy, Office of Science, Basic Energy Sciences. This work of S.K. was developed within the scope of the project CICECO-Aveiro Institute of Materials, UIDB/50011/2020(DOI 10.54499/UIDB/50011/2020), UIDP/50011/2020 (DOI 10.54499/UIDP/50011/2020) & LA/P/0006/2020 (DOI10.54499/LA/P/0006/2020), financed by national funds through the FCT/MCTES (PIDDAC). Part of S.K. work was funded by national funds (OE), through FCT - Fundação para a Ciência e a Tecnologia, I.P., in the scope of the framework contract foreseen in the numbers 4, 5, and 6 of article 23, of the Decree-Law 57/2016, of 29 August, changed by Law 57/2017, of 19 July. Y.M.V. and S.K. acknowledges support from the Horizon Europe Framework Programme (HORIZON-TMA-MSCA-SE), project № 101131229, Piezoelectricity in 2D-materials: materials, modeling, and applications (PIEZO 2D). Numerical results were visualized in Mathematica 14.0 [71].

**Authors' contribution.** A.N.M. and S.V.K. generated the research idea, analyzed results and wrote the manuscript draft. A.N.M. formulated the problem and performed analytical calculations. E.A.E. wrote the codes and prepared figures. S.V.K., Y.M.V. and D.R.E. worked on the results explanation and manuscript improvement. All co-authors discussed the results.

# Supplemental Material to

# "Ferri-ionic Coupling in CuInP$_2$S$_6$ Nanoflakes: Polarization States and Controllable Negative Capacitance"

Anna N. Morozovska[1,*], Sergei V. Kalinin[2,†], Eugene. A. Eliseev[3], Svitlana Kopyl[4], Yulian M. Vysochanskii[5], and Dean R. Evans[6,‡]

[1] Institute of Physics, National Academy of Sciences of Ukraine, 46, pr. Nauky, 03028 Kyiv, Ukraine

[2] Department of Materials Science and Engineering, University of Tennessee, Knoxville, TN, 37996, USA

[3] Frantsevich Institute for Problems in Materials Science, National Academy of Sciences of Ukraine, Omeliana Pritsaka str., 3, Kyiv, 03142, Ukraine

[4] Department of Physics & CICECO – Aveiro Institute of Materials, Campus Universitario de Santiago, 3810-193 Aveiro, Portugal

[5] Institute of Solid-State Physics and Chemistry, Uzhhorod University, 88000 Uzhhorod, Ukraine

[6] Air Force Research Laboratory, Materials and Manufacturing Directorate, Wright-Patterson Air Force Base, Ohio, 45433, USA

## APPENDIX A. Basic equations and approximations
### A1. The electroneutrality condition for surface charges

The electroneutrality condition equivalent to the total charge absence,

$$\sigma_{p0} + \sigma_{n0} = 0, \tag{A.1}$$

should be valid for the pairwise formation of negative and positive surface charges at $\phi = 0$, and the condition imposes limitations on the charge density parameters in Eqs.(4), namely:

$$\frac{Z_p C_p}{Z_n C_n} = -\frac{1 + g_p \exp\left(\frac{\Delta G_p^0}{k_B T}\right)}{1 + g_n \exp\left(\frac{\Delta G_n^0}{k_B T}\right)}. \tag{A.2}$$

To fulfil this condition (A.2), we consider the pairwise formation of negative and positive surface charges, when the charges have opposite signs, $Z_p = -Z_n = Z$, and their concentrations are equal $C_p = C_n = C$. Under the condition (A.2), the following relation between the prefactors and formation energies,

---

[*] corresponding author, e-mail: anna.n.morozovska@gmail.com
[†] corresponding author, e-mail: sergei2@utk.edu
[‡] corresponding author, e-mail: dean.evans@afrl.af.mil

$$\Delta G_p^0 - \Delta G_n^0 = k_B T ln\left(\frac{g_n}{g_p}\right), \tag{A.3}$$

is valid. If the prefactors are equal, $g_p = g_n = g$, then the formation energies are also equal, $\Delta G_p^0 = \Delta G_n^0 = \Delta G$. A violation of conditions (A.1)-(A.3) can take place far from equilibrium, e.g., under electrochemical reactions considered elsewhere [1, 2].

### A2. LGD free energy of CIPS nanoflakes

The polar and antipolar order parameters in CIPS originate from the antiparallel dipolar displacements of $Cu^+$ and $In^{3+}$ cations. Quantitatively the situation is much more complex, because one should consider one polar and 3 possible antipolar orderings in the $CuMP_2(S,Se)_6$ (M = In or Cr) structure [3, 43]. The polar ($P_i$) and three antipolar order parameters ($A_i$, $B_i$, and $\tilde{A}_i$) are related to the four atomic displacements $U_i^{(m)}$ of polar-active groups in the $CuMP_2(S,Se)_6$ structure as [3]:

$$P_i = \frac{\rho}{2}\left(U_i^{(1)} + U_i^{(2)} + U_i^{(3)} + U_i^{(4)}\right), \quad A_i = \frac{\rho}{2}\left(U_i^{(1)} - U_i^{(2)} - U_i^{(3)} + U_i^{(4)}\right), \tag{A.4a}$$

$$B_i = \frac{\rho}{2}\left(U_i^{(1)} - U_i^{(2)} + U_i^{(3)} - U_i^{(4)}\right), \quad \tilde{A}_i = \frac{\rho}{2}\left(U_i^{(1)} + U_i^{(2)} - U_i^{(3)} - U_i^{(4)}\right). \tag{A.4b}$$

Here $\rho \cong \frac{Q^*}{V}$ is a dimensionality factor, which is proportional to the effective Born charge $Q^*$ divided by the unit cell volume $V$, and $i = 1, 2, 3$.

In the most common cases, two out of four combinations of atomic displacements can be assumed to be zero, e.g., $\tilde{A}_i = B_i = 0$ (or $A_i = B_i = 0$). Corresponding displacements $U_i^{(m)}$ can be expressed via a nonzero polar parameter $P_i$ and an antipolar parameter $A_i$ as $U_i^{(1)} = U_i^{(4)} = \frac{P_i + A_i}{2\rho}$ and $U_i^{(2)} = U_i^{(3)} = \frac{P_i - A_i}{2\rho}$. The displacements are equal in the homogeneous ferroelectric (FE) phase, $U_i^{(1)} = U_i^{(2)} = U_i^{(3)} = U_i^{(4)} = \frac{P_i}{2\rho}$. In the case of the ferrielectric phases FI1 and FI2, the displacements are not equal, but have the same sign. The displacements $U_i^{(1)} = -U_i^{(2)} = -U_i^{(3)} = U_i^{(4)} = \frac{A_i}{2\rho}$, or $U_i^{(1)} = U_i^{(2)} = -U_i^{(3)} = -U_i^{(4)} = \frac{A_i}{2\rho}$, or $U_i^{(1)} = U_i^{(3)} = -U_i^{(2)} = -U_i^{(4)} = \frac{A_i}{2\rho}$ correspond to the three antiferroelectric states, AFE1, AFE2, and AFE3, which are predicted by the DFT calculations in Ref. [4].

The case $\tilde{A}_i = B_i = 0$, considered in this work, allows for a simplification of the description of $CuMP_2(S,Se)_6$ polar properties to a conventional LGD-type formalism. The case $A_i = B_i = 0$ can be considered in a similar way. A complete LGD thermodynamic potential describing ferrielectric CIPS with a first order phase transition contains even (2-nd, 4-th, and 6-th) powers of $P_i$ and $A_i$, as well as the biquadratic coupling between them. As shown in Ref. [5], the biquadratic coupling term $A_i^2 P_j^2$ induces a term proportional to $P^8$ in the LGD thermodynamic potential for $P_i$. Thus, the LGD free energy functional $G$ of a uniaxial ferroelectric includes a Landau energy – an expansion on 2-4-6-8 powers of the polarization component $P_3$, $G_{Landau}$; a polarization gradient energy, $G_{grad}$; an electrostatic energy, $G_{elect}$; an elastic, electrostriction, and flexoelectric contributions, $G_{elastic}$; and a surface energy, $G_S$. It has the form:

$$G = G_{Landau} + G_{grad} + G_{elect} + G_{elastic} + G_S, \qquad (A.5a)$$

where:

$$G_{Landau} = \int_{V_f} d^3r \left( \frac{\alpha}{2} P_3^2 + \frac{\beta}{4} P_3^4 + \frac{\gamma}{6} P_3^6 + \frac{\delta}{6} P_3^8 \right), \qquad (A.5b)$$

$$G_{grad} = \int_{V_C} d^3r \frac{g_{33kl}}{2} \frac{\partial P_3}{\partial x_k} \frac{\partial P_3}{\partial x_l}, \qquad (A.5c)$$

$$G_{elect} = -\int_{V_f} d^3r \left( P_i E_i + \frac{\varepsilon_0 \varepsilon_b}{2} E_i E_i \right), \qquad (A.5d)$$

$$G_{es} = -\int_V d^3r \left( \frac{s_{ijkl}}{2} \sigma_{ij} \sigma_{kl} + Q_{ij33} \sigma_{ij} P_3^2 + Z_{ij33} \sigma_{ij} P_3^4 + \frac{F_{ij3l}}{2} \left( \sigma_{ij} \frac{\partial P_3}{\partial x_l} - P_3 \frac{\partial \sigma_{ij}}{\partial x_l} \right) \right), \qquad (A.5e)$$

$$G_S = \frac{1}{2} \int_S d^2r\, a_{33}^{(S)} P_3^2. \qquad (A.5f)$$

Here $V = hS$ ($S$ = surface area, $h$ = thickness) is the nanoflake volume. The coefficient $\alpha$ linearly depends on temperature $T$, $\alpha(T) = \alpha_T(T - T_C)$, where $\alpha_T$ is the inverse Curie-Weiss constant and $T_C$ is the Curie temperature of a bulk CIPS. All other coefficients in Eq.(A.5a) are regarded as being temperature-independent. The coefficient $\beta$ is positive if the ferroelectric material undergoes a second order transition to the paraelectric phase and negative otherwise; and the coefficient $\gamma \geq 0$. The values $g_{ijkl}$ are the components of the gradient coefficients' tensor, which matrix should be positively defined. In Eq.(A.5d), $\sigma_{ij}$ is the stress tensor, $s_{ijkl}$ is the elastic compliances tensor, $Q_{ijkl}$ is the electrostriction tensor, and $F_{ijkl}$ is the flexoelectric tensor. The values $T_C$, $\alpha_T$, $\beta$, $\gamma$, $\delta$, $Q_{ijkl}$, and $Z_{ijkl}$ in **Table AI** were defined for CIPS from the fitting of experimentally observed temperature dependence of dielectric permittivity [6, 7, 8], spontaneous polarization [9], and lattice constants [10] for hydrostatic and uniaxial pressures. The elastic compliances $s_{ij}$ were estimated from ultrasound velocity measurements [11, 12, 13].

The electrostatic energy, $G_{elect}$, includes the relative background permittivity, $\varepsilon_b$. The inclusion of the relative background permittivity is a common rule for a Landau-type description of the dielectric properties of various ferroics with a soft polar optical mode (i.e., for the case of ferroelectrics, ferrielectrics, and paraelectrics); this is in a full agreement with multiple experimental results proving the existence of a constant "background" permittivity far from the Curie temperature $T_C$ (see e.g., Refs. [14, 15]).

According to experimental results [14, 15], the total polarization of the ferroic, $P_k^t$, includes both a ferro- (or ferri-) electric contribution $P_k$, originating from the polar soft mode, and a "background" contribution, $P_k^b$,

$$P_k^t(\vec{E}) = P_k(\vec{E}) + P_k^b(\vec{E}), \quad P_k^b(\vec{E}) = \varepsilon_0(\varepsilon_{km}^b - \delta_{km})E_m. \tag{A.5g}$$

Here $P_k(\vec{E})$ are the ferro- (or ferri-) electric polarization components included in the Landau-type free energy $G_{Landau}$. The "background" contribution, $P_k^b$, depends linearly on the electric field $\vec{E}$, where $\varepsilon_0$ is the dielectric permittivity of vacuum, $\delta_{km}$ is a Kroneker symbol, and $\varepsilon_{km}^b$ is an electric field-independent background permittivity [16].

**Table AI.** LGD parameters for a CuInP$_2$S$_6$ nanoflake

| Coefficient | Units | Numerical value |
|---|---|---|
| $\varepsilon_b$ | dimensionless | 9 |
| $\alpha_T$ | C$^{-2}$·m J/K | 1.64067×10$^7$ |
| $T_C$ | K | 292.67 |
| $\beta$ | C$^{-4}$·m$^5$J | 3.148×10$^{12}$ |
| $\gamma$ | C$^{-6}$·m$^9$J | −1.0776×10$^{16}$ |
| $\delta$ | C$^{-8}$·m$^{13}$J | 7.6318×10$^{18}$ |
| $Q_{i3}$ | C$^{-2}$·m$^4$ | $Q_{13} = 1.70136 - 0.00363\,T$, $Q_{23} = 1.13424 - 0.00242\,T$, $Q_{33} = -5.622 + 0.0105\,T$ |
| $Z_{i33}$ | C$^{-4}$·m$^8$ | $Z_{133} = -2059.65 + 0.8\,T$, $Z_{233} = -1211.26 + 0.45\,T$, $Z_{333} = 1381.37 - 12\,T$ |
| $W_{ij3}$ | C$^{-2}$·m$^4$ Pa$^{-1}$ | $W_{113} \approx W_{223} \approx W_{333} \cong -2\times10^{-12}$ |
| $s_{ij}$ | Pa$^{-1}$ | $s_{11} = 1.092\times10^{-11}$, $s_{12} = -0.311\times10^{-11}$, $s_{22} = 1.074\times10^{-11}$ |
| $g_{33ij}$ | J m$^3$/C$^2$ | $g \cong 2\times10^{-9}$ |
| $\lambda$ | nm | 10 |
| $R$ | µm | 0.1 - 5 |
| $h$ | nm | 5 - 100 |

### A3. Calculations of free charge at the electrodes

Let us consider a capacitor of thickness $h + d$ filled with a polarized ferroelectric film of thickness $h$ sandwiched between the bottom electrode and the dielectric layer of thickness $d$, as shown in **Fig. 1**.

The connection between the electric displacement $\vec{D}$ and the electric field, $\vec{E} = -\nabla\varphi$, in the ferroelectric ($f$) and the dielectric ($d$) layer are:

$$\vec{D}^f = \varepsilon_0\varepsilon_b\vec{E}^f + \vec{P}, \qquad \vec{D}^d = \varepsilon_0\varepsilon_d\vec{E}^l, \tag{A.6}$$

where $\varepsilon_0$ is a universal dielectric constant, $\varepsilon_b$ is a relative background permittivity of the ferroelectric, and $\varepsilon_d$ is relative dielectric permittivity of the dielectric layer. In the case when all variables depend on the z-coordinate and the ferroelectric polarization $\vec{P}$ is directed along the polar axis $z$, $\vec{P} = (0,0,P)$, we obtain the following electrostatic equations for the electric potential $\varphi_f$ in the ferroelectric film and for the electric potential $\varphi_d$ in dielectric the layer:

$$\varepsilon_0 \varepsilon_b \frac{\partial^2 \varphi_f}{\partial z^2} = \frac{\partial P}{\partial z}, \qquad \varepsilon_0 \varepsilon_s \frac{\partial^2 \varphi_d(z)}{\partial z^2} = 0. \tag{A.7}$$

The tangential components of the electric field are homogenous at the interfaces, and the normal components of the displacement differ by the value of the surface charge densities $\sigma_n$ and $\sigma_p$ at the interface $z = h$. Taking this into account, the boundary conditions are as follows:

$$\varphi_d(h+d) = U, \tag{A.8a}$$

$$\varphi_f(h) = \varphi_d(h), \tag{A.8b}$$

$$D_z^d(h) - D_z^f(h) = \sigma_n + \sigma_p, \tag{A.8c}$$

$$\varphi_f(0) = 0. \tag{A.8d}$$

Using the superposition principle, we can consider the general solution of Eqs.(A.7) in the following form:

$$\varphi_d(z) = -\frac{D_d(z-h-d)}{\varepsilon_0 \varepsilon_d} + U, \tag{A.9a}$$

$$\varphi_f(z) = -\frac{D_f z}{\varepsilon_0 \varepsilon_b} + \frac{1}{\varepsilon_0 \varepsilon_b} \int_0^z P(\tilde{z}) d\tilde{z}. \tag{A.9b}$$

Here the conditions (A.8a) and (A.8d) are already satisfied. After the substitution of the expressions (A.9) into the rest of the boundary conditions (A.8) and using expressions (A.6), one obtains the following system of equations for the unknown displacements $D_f$ and $D_d$:

$$\begin{cases} U + \frac{h}{\varepsilon_0 \varepsilon_b} D_f - \frac{\int_0^h P(\tilde{z}) d\tilde{z}}{\varepsilon_0 \varepsilon_b} + \frac{d}{\varepsilon_0 \varepsilon_d} D_d = 0, \\ D_f + \sigma - D_d = 0. \end{cases} \tag{A.10}$$

Here $\sigma_n + \sigma_p = \sigma$. The elementary transformations of Eq.(A.10) lead to the solution for the electric displacement $D_f$, electric field $E_f$, and electrostatic potential $\varphi_f$ inside the ferroelectric are:

$$D_f = -\frac{d\sigma \varepsilon_b - \varepsilon_d \int_0^h P(\tilde{z}) d\tilde{z} + U \varepsilon_0 \varepsilon_b \varepsilon_d}{d\varepsilon_b + h\varepsilon_d}, \tag{A.11a}$$

$$E_f = -\frac{P(z)}{\varepsilon_0 \varepsilon_b} + \frac{\varepsilon_d}{\varepsilon_0 \varepsilon_b (d\varepsilon_b + h\varepsilon_d)} \int_0^h P(\tilde{z}) d\tilde{z} - \frac{d\sigma}{(d\varepsilon_b + h\varepsilon_d)\varepsilon_0} - \frac{U \varepsilon_d}{d\varepsilon_b + h\varepsilon_d}, \tag{A.11b}$$

$$\varphi_f(z) = \frac{1}{\varepsilon_0 \varepsilon_b} \int_0^z P(\tilde{z}) d\tilde{z} - z \left\{ \frac{\varepsilon_d}{(d\varepsilon_b + h\varepsilon_d)\varepsilon_0 \varepsilon_b} \int_0^h P(\tilde{z}) d\tilde{z} - \frac{d\sigma}{(d\varepsilon_b + h\varepsilon_d)\varepsilon_0} - \frac{U \varepsilon_d}{d\varepsilon_b + h\varepsilon_d} \right\}. \tag{A.11c}$$

For the case of constant polarization, $P(\tilde{z}) \equiv P$, the field inside ferroelectric is

$$E_f = -\frac{d P}{(d\varepsilon_b + h\varepsilon_d)\varepsilon_0} - \frac{d \sigma}{(d\varepsilon_b + h\varepsilon_d)\varepsilon_0} - \frac{U \varepsilon_d}{d\varepsilon_b + h\varepsilon_d}. \tag{A.12}$$

The electric displacement $D_d$, electric field $E_d$, and electrostatic potential $\varphi_d$ inside the dielectric layer are given by expressions:

$$D_d = \frac{\varepsilon_d}{d\varepsilon_b + h\varepsilon_d}\left(h\,\sigma + \int_0^h P(\tilde{z})d\tilde{z} - U\varepsilon_0\varepsilon_b\right), \tag{A.13a}$$

$$E_d = \frac{h\,\sigma + \int_0^h P(\tilde{z})d\tilde{z} - U\varepsilon_0\varepsilon_b}{(d\varepsilon_b + h\varepsilon_d)\varepsilon_0}, \tag{A.13b}$$

$$\varphi_d(z) = U - (z - d - h)\frac{h\,\sigma + \int_0^h P(\tilde{z})d\tilde{z} - U\varepsilon_0\varepsilon_b}{(d\varepsilon_b + h\varepsilon_d)\varepsilon_0}. \tag{A.13c}$$

The interface potential has the form:

$$\varphi_f(h) = \frac{h\,d(\bar{P} + \sigma)}{(d\varepsilon_b + h\varepsilon_d)\varepsilon_0} + \frac{U\,h\,\varepsilon_d}{d\varepsilon_b + h\varepsilon_d}. \tag{A.14}$$

Charges at upper and lower electrodes, $q_1$ and $q_2$, are equal to

$$q_1 = -D_z^d(h+d) = -\frac{h\sigma\varepsilon_d}{d\varepsilon_b + h\varepsilon_d} - \frac{h\bar{P}\varepsilon_d}{d\varepsilon_b + h\varepsilon_d} + \frac{U\varepsilon_0\varepsilon_b\varepsilon_d}{d\varepsilon_b + h\varepsilon_d}, \tag{A.15a}$$

$$q_2 = +D_z^f(0) = -\frac{d\sigma\varepsilon_b}{d\varepsilon_b + h\varepsilon_d} + \frac{h\bar{P}\varepsilon_d}{d\varepsilon_b + h\varepsilon_d} - \frac{U\varepsilon_0\varepsilon_b\varepsilon_d}{d\varepsilon_b + h\varepsilon_d}. \tag{A.15b}$$

The discharge $q$ is equal to:

$$q(U) = \frac{q_1 - q_2}{2} = \frac{-h\varepsilon_d + d\varepsilon_b}{2(d\varepsilon_b + h\varepsilon_d)}\sigma - \frac{h\varepsilon_d}{d\varepsilon_b + h\varepsilon_d}\bar{P} + \frac{\varepsilon_0\varepsilon_b\varepsilon_d}{d\varepsilon_b + h\varepsilon_d}U. \tag{A.16}$$

Approximate equalities in Eqs.(A.15), valid for $d \ll h$, are:

$$q_1 = -\frac{h\varepsilon_d}{d\varepsilon_b + h\varepsilon_d}\sigma - \frac{h\varepsilon_d}{d\varepsilon_b + h\varepsilon_d}\bar{P} + \frac{\varepsilon_0\varepsilon_b\varepsilon_d}{d\varepsilon_b + h\varepsilon_d}U \approx -\sigma - \bar{P} + \frac{\varepsilon_0\varepsilon_b}{h}U, \tag{A.17a}$$

$$q_2 = -\frac{d\varepsilon_b}{d\varepsilon_b + h\varepsilon_d}\sigma + \frac{h\varepsilon_d}{d\varepsilon_b + h\varepsilon_d}\bar{P} - \frac{\varepsilon_0\varepsilon_b\varepsilon_d}{d\varepsilon_b + h\varepsilon_d}U \approx +\bar{P} - \frac{\varepsilon_0\varepsilon_b}{h}U. \tag{A.17b}$$

Note that the sum $q_2 + q_1 + \sigma$ is always equal to zero, as it follows from the system electroneutrality condition. Corresponding bound charges, surface free charges, and the first (i.e., nearest to the electrode surfaces) image charges for $U = 0$ are shown in the **Fig. A1** for the "up" (left side) and "down" (right side) directions of the CIPS spontaneous polarization.

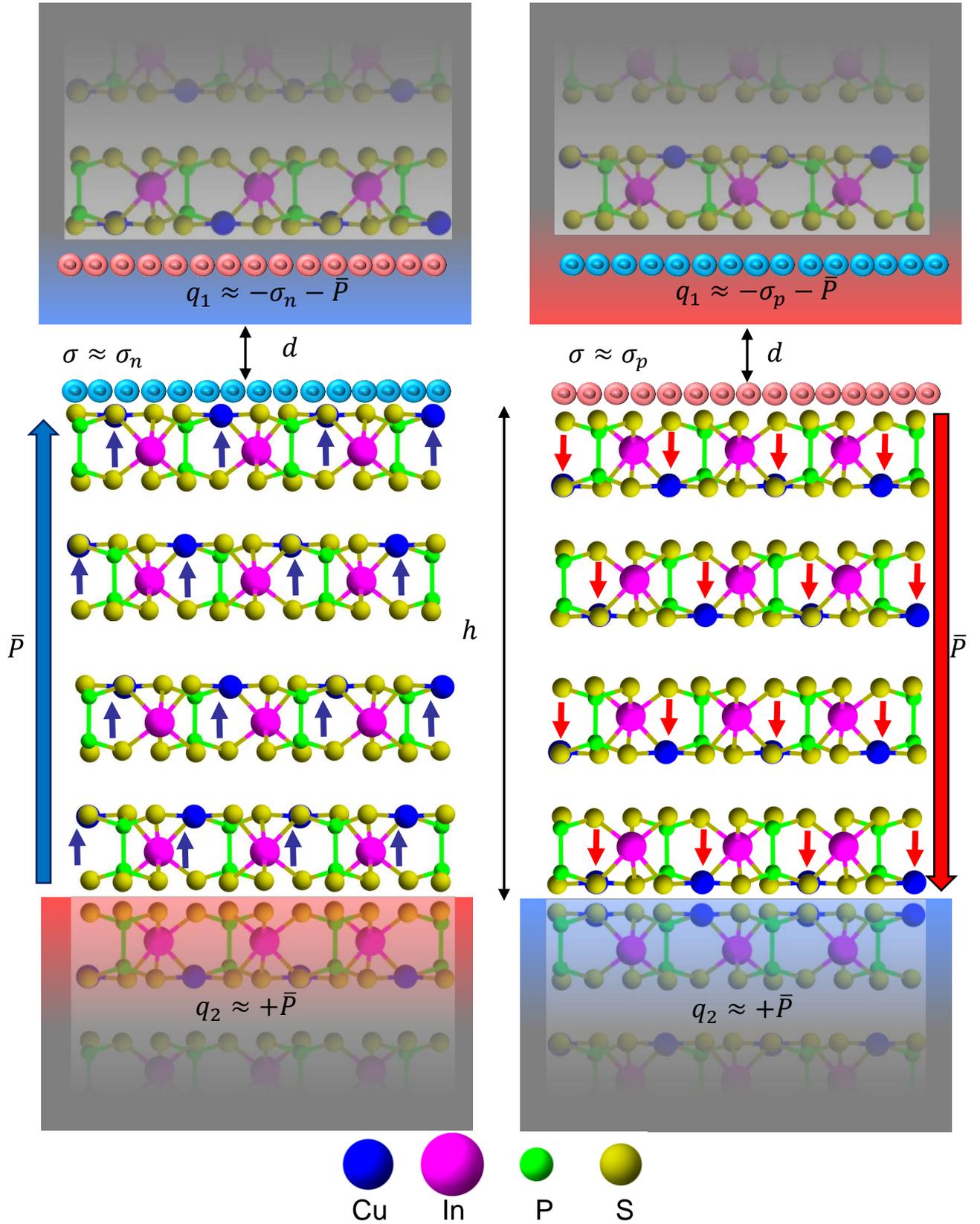

**FIGURE A1**. Schematic distribution of bound, free, and the first (i.e., nearest to the electrode surfaces) image charges in the considered structure "electrode - four-layer CIPS – surface charge – dielectric gap – electrode" for the "up" (left side) and "down" (right side) directions of the CIPS spontaneous polarization.

## A4. Polarization and charge hysteresis loops, maximal polarization, and stored charge in CIPS nanoflakes

Typical quasi-static hysteresis loops of the average polarization $\bar{P}(U)$, positive and negative surface charges, $\sigma_p(U)$ and $\sigma_n(U)$, and the stored charge $q(U)$ are shown in **Fig. A2** and **Fig. A3**, for $h = 10$ nm and 100 nm, respectively. The dependences of saturated polarization $\bar{P}(U_{max})$ and maximal stored charge $q(U_{max})$ on $C$ and $u_m$ are shown in **Fig. A4** for $h =$10 nm and 100 nm.

Hysteresis loops for 10-nm nanoflakes covered with surface charges in small concentrations, $C \leq 0.03$ nm$^{-2}$, are shown in **Fig.A2** (left side). A change of the mismatch strain from compressive strain (top row, $u_m = -0.25\%$) to tensile strain (bottom row, $u_m = +0.6\%$) leads to the following gradual transformation of $\bar{P}(U)$. A rectangular-shaped single hysteresis loop ($u_m = -0.25\%$) transforms into a double (or very strongly pinched) loop ($u_m = -0.1\%$), and then into a hysteresis-less paraelectric curve ($u_m = 0\%$) that persists until larger tensile strains ($u_m = +0.6\%$) are applied.

We can see a growing deviation between the $\bar{P}(U)$ and $q(U)$ loops with increasing $\sigma_p(U)$ and $\sigma_n(U)$; this is more noticeable at larger voltages (compare black and green traces in **Fig. A2** (left side)). The values of $\sigma_p(U)$ and $\sigma_n(U)$ are relatively small and increase slightly with an increase in $C$.

Note, for the case of a small surface change concentration (0.01 nm$^{-2} \leq C \leq 0.03$ nm$^{-2}$), the coercive voltage $U_c$ for $\bar{P}(U)$ and $q(U)$ are significantly larger for a compressive strain compared to tensile strain. Namely, $U_c$ is very small or nearly zero for almost all $u_m$ except the extremes of compressive ($u_m = -0.25\%$) and tensile ($u_m = +0.6\%$) strains. There is a weak dependence of $U_c$ on surface charge in the case of a compressive strain, and a much stronger dependence for the case of a tensile strain.

Hysteresis loops for 10-nm nanoflakes covered with surface charges in higher concentrations, $C > 0.03$ nm$^{-2}$, are shown in **Fig. A2** (right side). A change of $u_m$ from $-0.25$ % (top row) to $+0.6\%$ (bottom row) leads to the following gradual transformation of $\bar{P}(U)$. A wide rectangular-shaped single hysteresis loop becomes narrower, and then transforms into a slightly pinched loop, a strongly pinched loop, a double loop, a hysteresis-less paraelectric curve. The hysteresis-less curves transform back to narrow single hysteresis loops with relatively weaker polarizations; the thin loops become wider and acquire a pronounced rectangular-like shape with a further increase of $u_m$ above 0.5 %, which is shown in the bottom row**.** An increase of $C$ supports the opening of single hysteresis loops of $\bar{P}$ and $q$ from pinched or double loops, and even from PE curves (which transforms into single narrow loops).

The coercive voltage $U_c$ for $\bar{P}$ and $q$ strongly increases (up to several times) with an increase in $C$ from 0.1 nm$^{-2}$ to 1 nm$^{-2}$ for the strongest compressive strains (see the top row in **Fig. A2**). For the strongest tensile strains, $U_c$ increases more than 10 times with the same increase in $C$ (see the bottom row in **Fig. A2**). Also evident is a strong change in the coercive voltage for $\bar{P}$ and $q$ (greater than a factor of ten) with a change in $u_m$ from compressive strains (e.g., $U_c \approx 2$ V for $u_m = -0.25$ % and $C = 1$ nm$^{-2}$) to tensile strains (for e.g., $U_c \approx 0.1$ V for $u_m = +0.25$ % and $C = 1$ nm$^{-2}$). Such strong changes are possible during single-domain polarization switching considered hereinafter.

Hysteresis loops for 100 nm nanoflakes covered with surface charges are shown in **Fig. A3**. For small concentrations of surface charge ($C \leq 0.03$ nm$^{-2}$) the stored charge loops are similar to the polarization loops; and a difference between them increases with increasing values of $C$ (compare the black and green loops on the left side of **Fig. A3**). The values of $\sigma_p(U)$ and $\sigma_n(U)$ are relatively small and increase with an increase in $C$.

For the case of 100-nm nanoflakes, the coercive voltage $U_c$ for $\bar{P}$ and $q$ is fairly independent of the surface charge concentration, $C$, up to 0.1 nm$^{-2}$ for compressive and zero mismatch strains. There is a minor dependence on $C$ at larger concentrations (0.3 – 1 nm$^{-2}$), which is greater for the largest strains $u_m$, although the difference is two times less compared to the case of smaller concentrations. Under a tensile strain, the same trend is observed, but with a greater dependence on $C$ for the largest tensile strains (nearly a factor of 10). For the case of 10-nm nanoflakes, the difference in $U_c$ is much greater; at large compressive strains there is a factor of 5 increase between the smallest and largest surface charge concentrations, and at the largest tensile strains this increases to a factor of 80. In comparing this effect between the two different thicknesses studied, one can conclude that the $U_c$ of thinner nanoflakes reveals a much stronger dependence on the surface charge concentration and mismatch strain, particularly for the tensile strains. The physical origin of this dependence is due to the larger surface to volume ratio in comparison to thick flakes. The ratio is proportional to the flake thickness $h$. According to Eq.(2a), the size effect of acting electric field scales as $\frac{d}{h}$ for $h \gg d$, and thus we can expect a $1/h$ scaling law in the surface charge impact on the flake properties.

Note, that the 10 times increase in the coercive voltage $U_c$ of the single loops for the 100-nm nanoflakes in comparison with the 10-nm nanoflakes (compare, e.g., the first columns in **Fig. A3** and **Fig. A2**) corresponds to the same coercive field $E_c$, because the proportionality law $E_c \cong \frac{U_c}{h}$ is valid for the single-domain polarization switching. From Eq.(1), the coercive field $E_c$ is proportional to $\left[\frac{d\varepsilon_d}{\varepsilon_0(d\varepsilon_b+h\varepsilon_d)} - \tilde{\alpha}(T)\right]^{3/2}$ for small $C$, i.e., $E_c$ decreases with $h$ increase. The proportionality relation $U_c \cong E_c h \sim h \left[\frac{d\varepsilon_d}{\varepsilon_0(d\varepsilon_b+h\varepsilon_d)} - \tilde{\alpha}(T)\right]^{3/2}$ explains the difference in coercive voltages for thicker and thinner nanoflakes. However, the domain formation in weakly screened nanoflakes (i.e., for a small concentration $C$ of surface charges) can easily break the proportionality law.

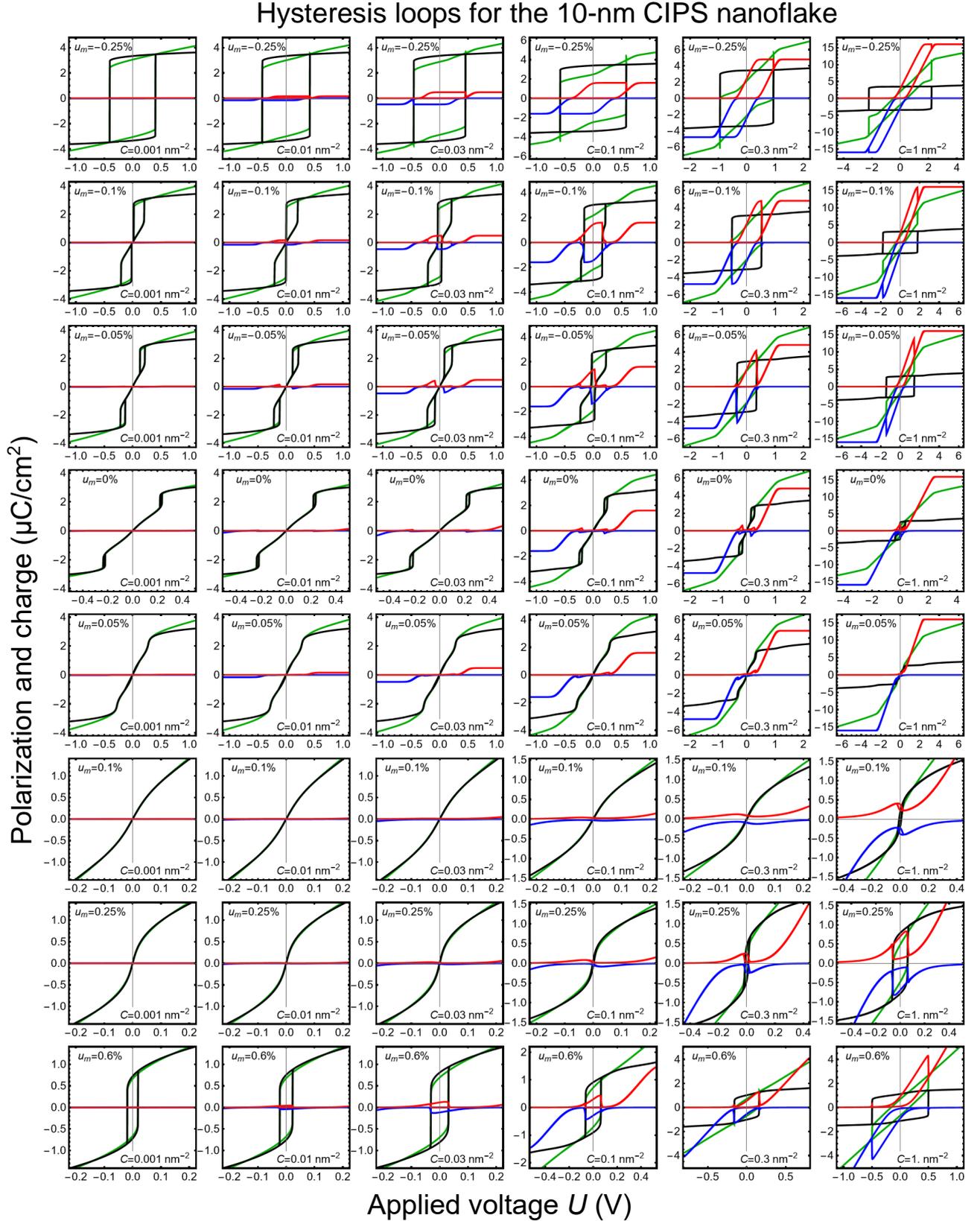

**FIGURE A2.** The quasi-static voltage dependences of the average polarization $\bar{P}(U)$ (black curves), positive (red

curses) and negative (blue curves) surface charges, $\sigma_p(U)$ and $\sigma_n(U)$, and stored charge $q(U)$ (green curves), calculated for 10-nm CIPS nanoflakes with different concentrations of surface charges $C$, varying from $10^{-3}$ nm$^{-2}$ to 1 nm$^{-2}$, and different degrees of mismatch strain $u_m$, ranging from -0.25% to +0.6%. Corresponding values of $C$ and $u_m$ are listed inside each plot. Other parameters: $h = 10$ nm, $R \geq 0.5$ μm, $d = 1$ nm, $\varepsilon_d = 10$, $\varepsilon_b = 9$, $g = 1$, $\Delta G = 0.1$ eV, $\tau_n = \tau_p = 10\tau_{Kh}$, and $T = 293$ K.

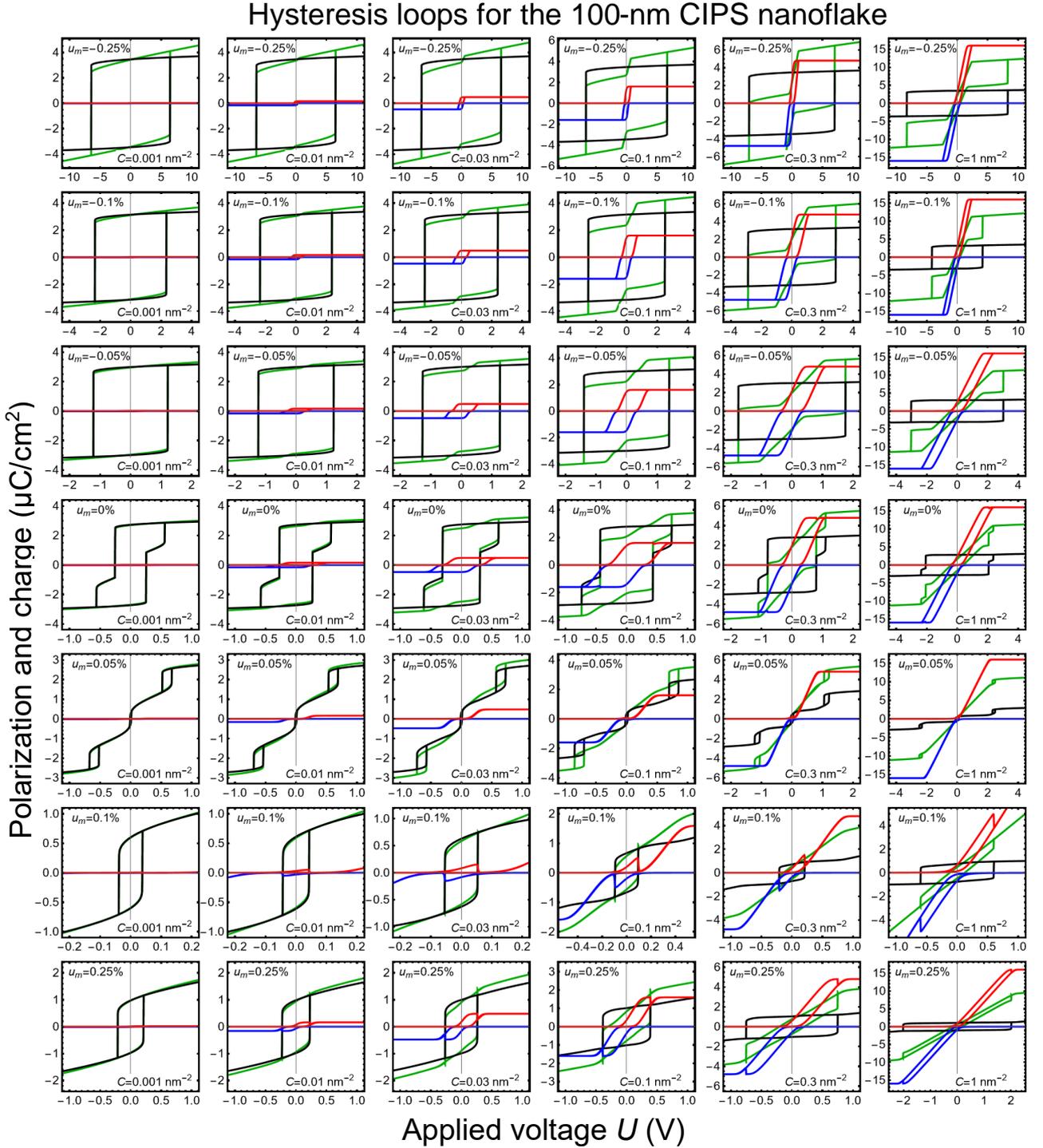

**FIGURE A3.** The quasi-static voltage dependences of the average polarization $\bar{P}(U)$ (black curves), positive (red

curves) and negative (blue curves) surface charges, $\sigma_p(U)$ and $\sigma_n(U)$, and stored charge $q(U)$ (green curves), calculated for 100-nm CIPS nanoflakes with different concentrations of surface charges $C$ varying from $10^{-3}$ nm$^{-2}$ to 1 nm$^{-2}$, and different degrees of mismatch strain $u_m$, ranging from -0.25% to +0.25%. Corresponding values of $C$ and $u_m$ are listed inside each plot. $h = 100$ nm and $R \geq 5$ μm. Other parameters are the same as in **Fig A2**.

The polarization value in saturation and the maximal stored free charge, $\bar{P}(U_{max})$ and $q(U_{max})$, calculated for the 10-nm CIPS nanoflake at $U_{max} = 2$ V are shown in **Figs. A4(a)** and **A4(b)**. The high external field (~2 V/nm) induces a relatively large electric polarization in the region of the PE phase, and increases the polarization in the ferrielectric and ferri-ionic regions; this leads to the complete disappearance of all phase boundaries in the diagrams **A4(a)** and **A4(b)** (in comparison with the clearly visible boundaries in the phase diagrams in **Fig. 4(a)** and **4(b)** in the main text). However, there are principal differences in the strain dependences of $\bar{P}(U_{max})$ and $q(U_{max})$. Namely, $\bar{P}(U_{max})$ reaches a maximum (~3.75 μC/cm$^2$) for the maximal compressive strain; it monotonically decreases with a strain change from compression to tension and is minimum (~1.3 μC/cm$^2$) for the maximal tensile strain. Also, $\bar{P}(U_{max})$ decreases monotonically with an increase in $C$; the blue region with the lowest $\bar{P}$ values is located in the top right corner of **Fig. A4(a)** corresponding to the greatest tensile strain and greatest charge concentration. The value of $Q(U_{max})$ weakly depends on the strain in comparison with $\bar{P}(U_{max})$ (see the color gradient in **Fig. A4(b)**, which is close to horizontal). Also, $q(U_{max})$ demonstrates the continuous transformation from a state with lower charge (~5 μC/cm$^2$, see the bottom part of **Fig. A4(b)**) to a state with higher charge (~10 μC/cm$^2$, see the top part of **Fig. A4(b)**) with an increase in $C$.

The polarization value in saturation and the maximal stored charge, $\bar{P}(U_{max})$ and $q(U_{max})$, calculated for the 100-nm CIPS flake at $U_{max} = 4$ V are shown in **Figs. A4(c)** and **A4(d)**, respectively. All main features of **Figs. A4(c)** and **A4(d)** are qualitatively comparable to those shown in **Figs. A4(a)** and **A4(b)**, respectively. Similar to the 10-nm flakes, high external fields induce an electric polarization in the thin region of the PE phase for the 100-nm flakes and increase the polarization in the ferrielectric and ferri-ionic regions. However, the values of $\bar{P}(U_{max})$ and $q(U_{max})$ saturate at higher voltages compared to those shown in **Figs. A4(a)** and **A4(b).**

The relatively weak dependence of the $\bar{P}(U_{max})$ on the thickness of CIPS nanoflakes contrasts with the much stronger thickness dependence of the $\bar{P}(0)$ (compare **Fig. 4(a)** and **4(c)** with **A4(a)** and **A4(c)**). The contrast can be explained by the dominant role of the external electric field, which is proportional to the applied voltage. The field determines the magnitude of the CIPS polarization in saturation. The dependence of the $q(U_{max})$ on the CIPS thickness is significant (compare **Fig. A4(b)** with **A4(d)**), because the expression (5) for $q(U)$ contains the voltage-dependent contribution, $-\frac{\varepsilon_0 \varepsilon_b \varepsilon_d}{d\varepsilon_b + h\varepsilon_d} U \approx -\frac{\varepsilon_0 \varepsilon_b}{h} U$, which also scales as $1/h$.

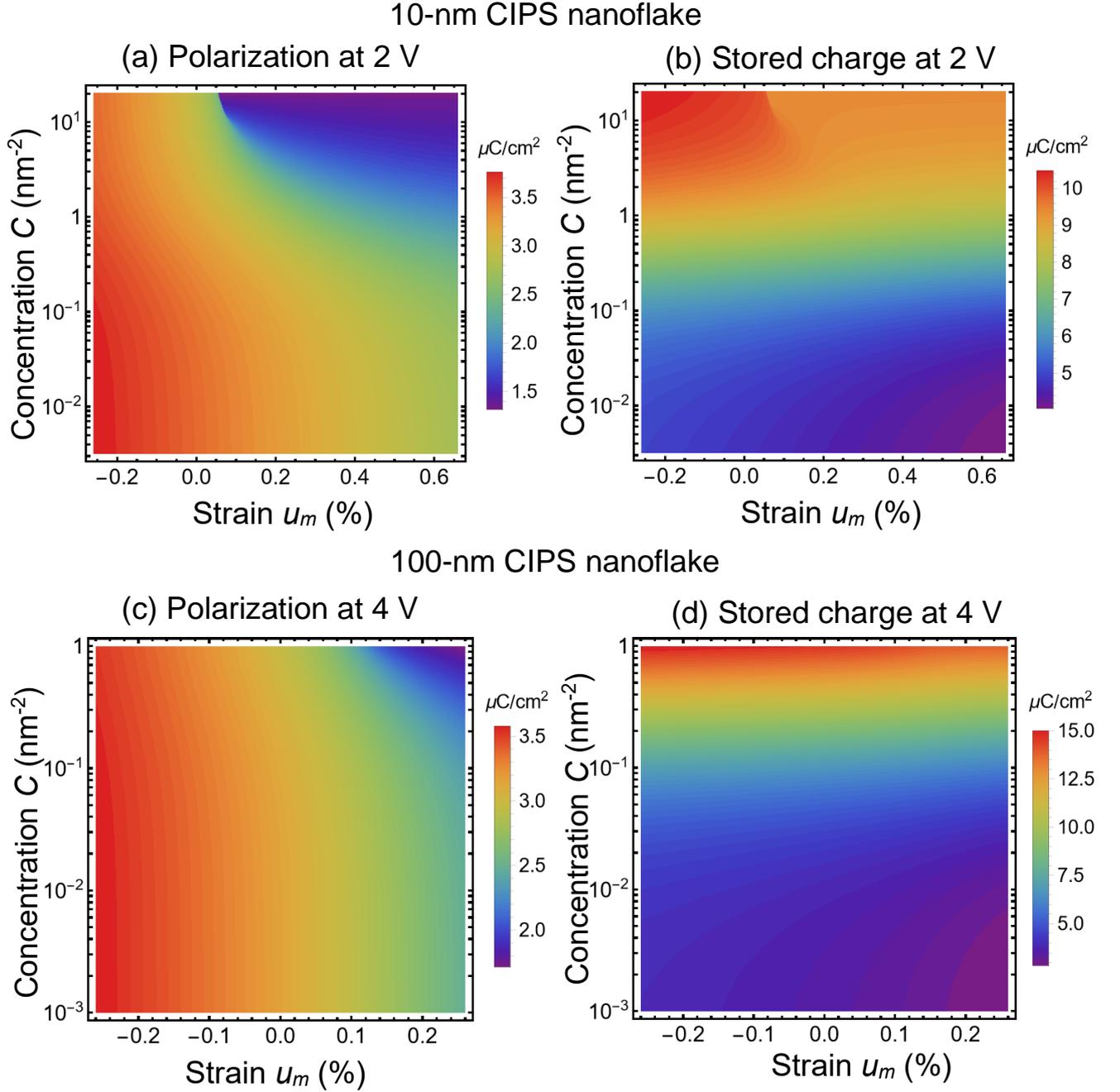

**FIGURE A4.** The dependences of the saturated polarization $\bar{P}(U_{max})$ **(a, c)** and maximal stored charge $q(U_{max})$ **(b, d)** on concentration of surface charges $C$ and mismatch strain $u_m$ for 10-nm **(a, b)** and 100-nm **(c, d)** CIPS nanoflakes. Here, $U_{max} = 2$ V for $h = 10$ nm and $U_{max} = 4$ V for $h = 100$ nm. Other parameters are the same as in **Fig 2**.

## A5. Analytical calculations of the negative capacitance effect

The charge of the reference SrTiO$_3$ capacitor is $Q_r = C_r U$, where the reference capacitance is $C_r = \frac{\varepsilon_0 \varepsilon_d}{d}$. The difference of the effective and reference capacitance is given by the expression:

$$\Delta C = C_r - \frac{dQ}{dU} = \frac{\varepsilon_0 \varepsilon_d}{d} + \frac{h\varepsilon_d - d\varepsilon_b}{d\varepsilon_b + h\varepsilon_d} \frac{d\sigma}{dU} + \frac{h\varepsilon_d}{d\varepsilon_b + h\varepsilon_d} \frac{d\bar{P}}{dU} - \frac{\varepsilon_0 \varepsilon_b \varepsilon_d}{d\varepsilon_b + h\varepsilon_d}. \tag{A.17}$$

The NC effect corresponds to the condition $\Delta C < 0$. The magnitude of $\bar{P}$ can be estimated from the Eqs.(1)-(2) in the main text:

$$\Gamma \frac{d}{dt} \bar{P} + \tilde{\alpha}(T)\bar{P} + \tilde{\beta}\bar{P}^3 + \tilde{\gamma}\bar{P}^5 + \tilde{\delta}\bar{P}^7 = -\frac{d}{\varepsilon_d h + \varepsilon_b d} \frac{\bar{P} + \sigma}{\varepsilon_0} - \frac{\varepsilon_d}{\varepsilon_d h + \varepsilon_b d} U, \tag{A.18}$$

where $\sigma = \sigma_p + \sigma_n$ is the total surface charge density.

Under the condition of a negligibly small contribution of the nonlinear polarization powers in Eq.(A.18), the terms $\tilde{\beta}\bar{P}^3 + \tilde{\gamma}\bar{P}^5 + \tilde{\delta}\bar{P}^7$ can be omitted, and the derivative $\frac{d\bar{P}}{dU}$ can be estimated as:

$$\frac{d\bar{P}}{dU} \approx -\frac{\varepsilon_d}{\varepsilon_d h + \varepsilon_b d} \left(1 + \frac{d}{\varepsilon_0} \frac{d\sigma}{dU}\right) \left[\tilde{\alpha} + \frac{d}{\varepsilon_0(\varepsilon_d h + \varepsilon_b d)}\right]^{-1}. \tag{A.19}$$

Under the validity of Eq.(A.19), the NC effect can be achieved under the conditions

$$\frac{h\varepsilon_d - d\varepsilon_b}{d\varepsilon_b + h\varepsilon_d} \frac{d\sigma}{dU} + \frac{h\varepsilon_d}{d\varepsilon_b + h\varepsilon_d} \frac{d\bar{P}}{dU} - \frac{\varepsilon_0 \varepsilon_b \varepsilon_d}{d\varepsilon_b + h\varepsilon_d} > -\frac{\varepsilon_0 \varepsilon_d}{d}, \qquad \tilde{\alpha} + \frac{d}{\varepsilon_0(\varepsilon_d h + \varepsilon_b d)} > 0, \tag{A.20}$$

where

$$\sigma[\phi] = eZC_n \left\{ \frac{1}{1+\exp\left(\frac{\Delta G + eZ\phi}{k_B T}\right)} - \frac{1}{1+\exp\left(\frac{\Delta G - eZ\phi}{k_B T}\right)} \right\} = eZC_n \frac{-2\exp\left(\frac{\Delta G}{k_B T}\right)\sinh\left(\frac{eZ\phi}{k_B T}\right)}{\left[1+\exp\left(\frac{\Delta G + eZ\phi}{k_B T}\right)\right]\left[1+\exp\left(\frac{\Delta G - eZ\phi}{k_B T}\right)\right]}, \tag{A.21a}$$

and

$$\phi = h\left(\frac{d}{\varepsilon_d h + \varepsilon_b d} \frac{\bar{P} + \sigma}{\varepsilon_0} + \frac{\varepsilon_d}{\varepsilon_d h + \varepsilon_b d} U\right). \tag{A.21b}$$

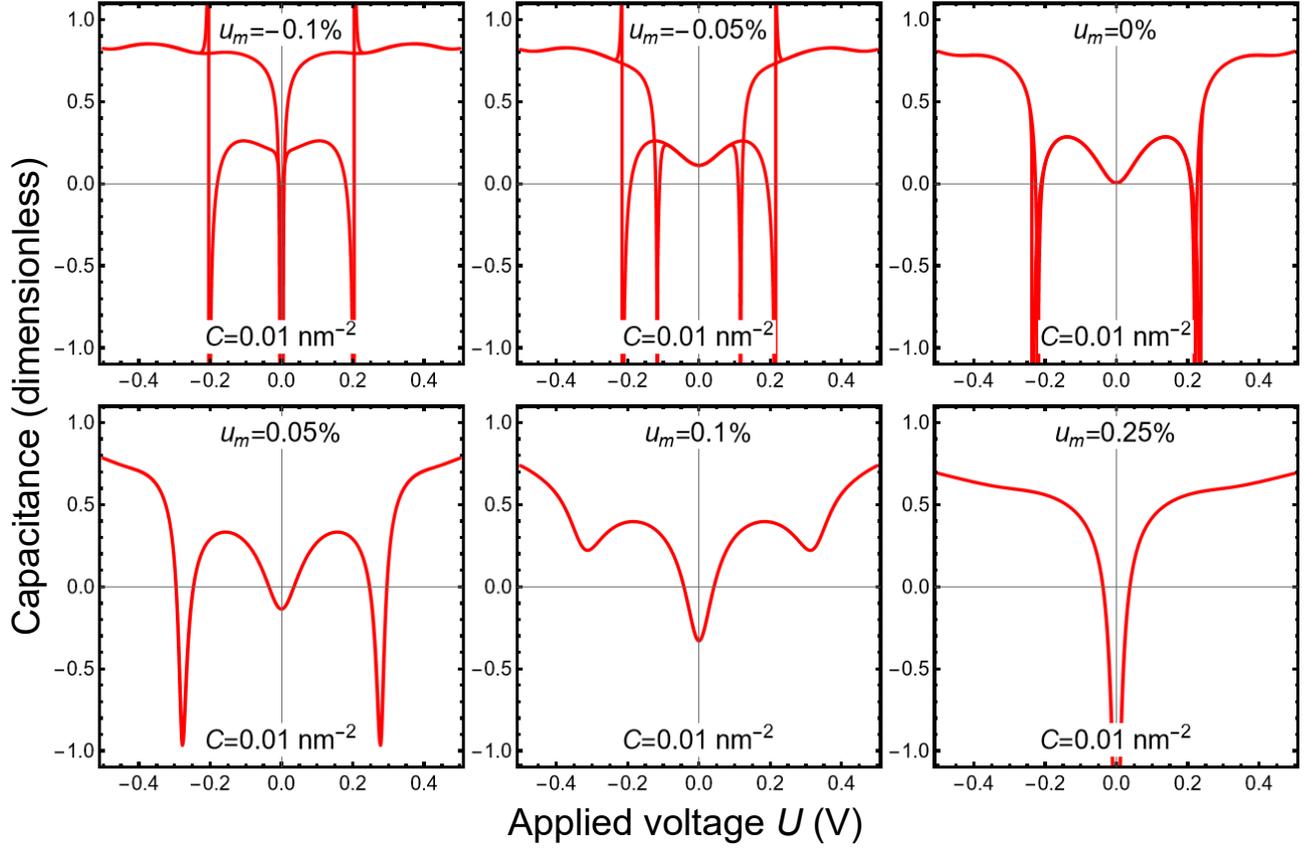

**FIGURE A5.** The quasi-static voltage dependences of the relative capacitance, $\Delta C/C_r$, calculated for different mismatch strains $u_m$, which vary from -0.1% to +0.3%, and a fixed value of concentration of surface charges, $C=10^{-3}$ nm$^{-2}$. Corresponding values of $C$ and $u_m$ are listed inside each plot. Nanoflake size: $h = 10$ nm and $R \geq 50$ nm; other parameters are the same as in **Fig A2**.

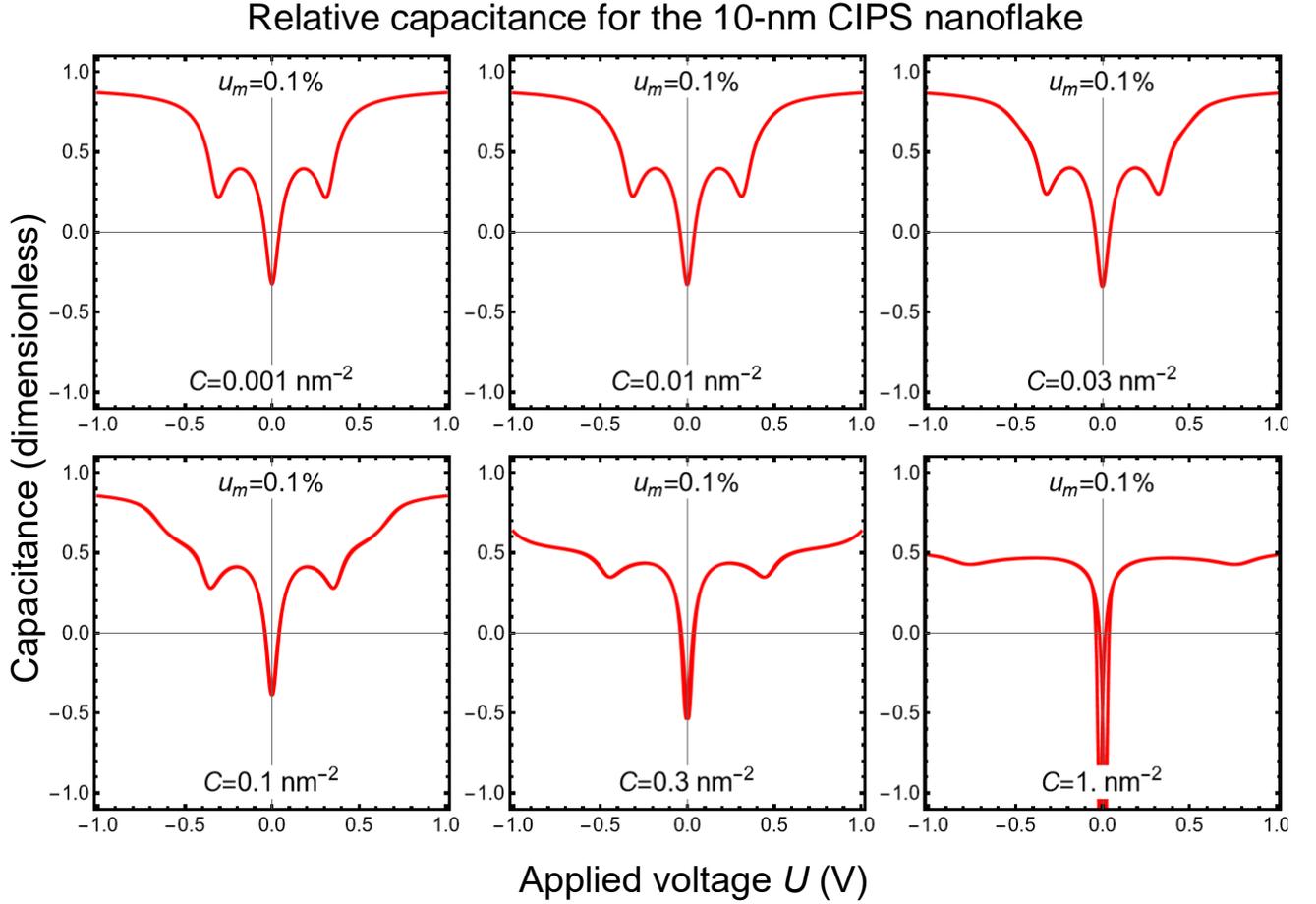

**FIGURE A6.** The quasi-static voltage dependences of the relative capacitance, $\frac{\Delta C}{C_r}$, calculated for different concentration surface charges $C$, which vary from $10^{-3}$ nm$^{-2}$ to 1 nm$^{-2}$, and a fixed value of mismatch strain $u_m$=+0.1%. Corresponding values of $C_n$ and $u_m$ are listed inside each plot. Nanoflake size: $h = 10$ nm and $R \geq 50$ nm; other parameters are the same as in **Fig A2**.